\pdfoutput=1
\documentclass[final,authoryear,3p,times]{elsarticle}

\usepackage{graphicx}
\usepackage{pstricks}
\usepackage{tikz}
\usepackage{subfig}
\graphicspath{{./figures/}}
\usepackage{amsmath,amssymb,amsfonts,amsthm,textcomp}
\usepackage{type1ec,type1cm,fix-cm,lmodern}
\usepackage[utf8]{inputenc}
\usepackage{color}
\usepackage{moreverb}
\usepackage{booktabs}
\usepackage{rotating}
\usepackage[section]{placeins}
\usepackage{xcolor}
\usepackage{tcolorbox}
\usepackage{stmaryrd}
\usepackage{tensor}
\usepackage{bm}
\usepackage{xstring}
\newcommand{\T}[2]{
  \ifcat\noexpand{#1}\relax % check if the argument is a control sequence
    \IfEqCase{#2}{
      {1}{{\bm{#1}}}
      {2}{{{\bm{#1}}}}
      {3}{{{{\bm{#1}}}}}
      {4}{{{{{\bm{#1}}}}}}
    }
  \else
    \IfEqCase{#2}{
      {1}{{\mathbf{#1}}}
      {2}{{{\mathbf{#1}}}}
      {3}{{{{\mathbf{#1}}}}}
      {4}{{{{{\mathbf{#1}}}}}}
    }
  \fi
  }
\newcommand{\varTd}[2]{
  \ifcat\noexpand{#1}\relax % check if the argument is a control sequence
    \IfEqCase{#2}{
      {1}{{\delta \bm{#1}}}
      {2}{{{\delta \bm{#1}}}}
      {3}{{{{\delta \bm{#1}}}}}
      {4}{{{{{\delta \bm{#1}}}}}}
    }
  \else
    \IfEqCase{#2}{
      {1}{{\delta \mathbf{#1}}}
      {2}{{{\delta \mathbf{#1}}}}
      {3}{{{{\delta \mathbf{#1}}}}}
      {4}{{{{{\delta \mathbf{#1}}}}}}
    }
  \fi
  }
\newcommand{\varTD}[2]{
  \ifcat\noexpand{#1}\relax % check if the argument is a control sequence
    \IfEqCase{#2}{
      {1}{{\Delta \bm{#1}}}
      {2}{{{\Delta \bm{#1}}}}
      {3}{{{{\Delta \bm{#1}}}}}
      {4}{{{{{\Delta \bm{#1}}}}}}
    }
  \else
    \IfEqCase{#2}{
      {1}{{\Delta \mathbf{#1}}}
      {2}{{{\Delta \mathbf{#1}}}}
      {3}{{{{\Delta \mathbf{#1}}}}}
      {4}{{{{{\Delta \mathbf{#1}}}}}}
    }
  \fi
  }

\newcommand{\eqdef}{\stackrel{\text{\scriptsize{def}}}{=}}

\newcommand{\jump}[1]{\llbracket {#1} \rrbracket}
\newcommand{\gradx}[1]{\nabla{#1}}

\newcommand{\divx}[1]{\nabla\cdot{#1}}

\newcommand{\diffo}[1]{\text{d}{#1}}

\newcommand{\diffn}[3]{\dfrac{\text{d}^{#3}{#1}}{\text{d}{#2}^{#3}}}
\newcommand{\partn}[3]{\dfrac{\partial^{#3}{#1}}{\partial{#2}^{#3}}}
\newcommand{\spart}[3]{\partial^{#3}_{#2}{#1}}
\newcommand{\matd}[1]{\dot{#1}}
\newtheoremstyle{slplain}% name
  {.5\baselineskip\@plus.2\baselineskip\@minus.2\baselineskip}% Space above
  {.5\baselineskip\@plus.2\baselineskip\@minus.2\baselineskip}% Space below
  {\slshape}% Body font
  {}%Indent amount (empty = no indent, \parindent = para indent)
  {\bfseries}%  Thm head font
  {.}%       Punctuation after thm head
  { }%      Space after thm head: " " = normal interword space;
  {}%       Thm head spec

\theoremstyle{slplain}
\newtheorem{rmk}{Remark}

\newcommand{\bt}[1]{{\color{blue}{#1}}}

\journal{arXiv}

\begin{document}

\begin{frontmatter}

\title{Energy Exchange Analysis in Droplet Dynamics via the Navier--Stokes--Cahn--Hilliard Model}

\author{LFR Espath\corref{cor1}\fnref{fn1}}
\ead{espath@gmail.com}

\author{AF Sarmiento\fnref{fn1}\fnref{fn2}}

\author{P Vignal\fnref{fn1}\fnref{fn3}}

\author{BON Varga\fnref{fn1}}

\author{AMA Cortes\fnref{fn1}}

\author{L Dalcin\fnref{fn1}\fnref{fn4}}

\author{VM Calo\fnref{fn1}\fnref{fn5}}

\cortext[cor1]{Corresponding author.}

\fntext[fn1]{Center for Numerical Porous Media, King Abdullah University of Science and Technology (KAUST), Thuwal, Saudi Arabia}

\fntext[fn2]{Applied Mathematics \& Computational Science, King Abdullah University of Science and Technology (KAUST), Thuwal, Saudi Arabia}

\fntext[fn3]{Material Science \& Engineering, King Abdullah University of Science and Technology (KAUST), Thuwal, Saudi Arabia}

\fntext[fn4]{National Scientific and Technical Research Council (CONICET), Santa Fe, Argentina}

\fntext[fn5]{Applied Geology Department, Western Australian School of Mines, Faculty of Science and Engineering, Curtin University, \\Perth, Western Australia, 6845, Australia}

\address{King Abdullah University of Science and Technology (KAUST), Thuwal, Saudi Arabia, \\
         Curtin University, Bentley, Perth, Western Australia, Australia}

\begin{abstract}
We develop the energy budget equation of the coupled Navier--Stokes--Cahn--Hilliard (NSCH) system. We use the NSCH equations to model the dynamics of liquid droplets in a liquid continuum. Buoyancy effects are accounted for through the Boussinesq assumption. We physically interpret each quantity involved in the energy exchange to further insight into the model. Highly resolved simulations involving density-driven flows and merging of droplets allow us to analyze these energy budgets. In particular, we focus on the energy exchanges when droplets merge, and describe flow features relevant to this phenomenon. By comparing our numerical simulations to analytical predictions and experimental results available in the literature, we conclude that modeling droplet dynamics within the framework of NSCH equations is a sensible approach worth further research.
\end{abstract}


\end{frontmatter}

%===============================================================================================================%

\section{Introduction}

Bubbles and droplets are ubiquitous in natural processes and technological applications. Engineers, mathematicians, and physicists have attempted to understand these phenomena over the last 200 years \citep{CLI05}. Studies can be traced back to Young \citep{YOU1805} and Laplace \citep{LAP1805}, who described the nonlinear partial differential equation governing the pressure jump developed across a curved interface between fluids. This model is known as the Young-Laplace equation. Processes that drive the study of the dynamics of bubbles and droplets include rainfall, air pollution, boiling, flotation, topological phase transition, and fermentation \citep{CLI05}. Also, there has been an intensive study of the coalescence of droplets \citep{KAV16} as well as the droplet impact on solid surfaces \citep{JOS16}.

Several methodologies have been developed to model pressure jumps and sharp gradients. Among them, sharp- and diffuse-interface methods are widely used to model phase segregation problems coupled with fluid dynamics. Sharp-interface models include the level-set \citep{SET99} and volume-of-fluid \citep{RID98} methods. Diffuse-interface models (DIM) have been used to model topological transitions in compressible \citep{AND98} and quasi-incompressible \citep{LOW98} flows under consistent thermodynamic premises. A wide range of phenomena is described via diffuse interfaces, going from material sciences \citep{LOG01} and fracture mechanics \citep{SPA07} to fluid dynamics \citep{KHA06,KHA07,GOM13}. A diffuse interface model circumvents several numerical difficulties \citep{YUE04} and its inherent thermodynamic consistency allows the method to incorporate rheology when the phase-field evolution is described by a free energy density \citep{EMM03}.

In this work we use the Cahn--Hilliard equation with the Ginzburg--Landau free energy defined in \citet{GOM08}. A derivation of the Cahn--Hilliard equation is presented by \citet{GUR96a} using a balance law for microforces. \citet{GUR96b} extended that work to couple phase segregation with hydrodynamics, providing the first derivation of the Navier--Stokes--Cahn--Hilliard system completely based on continuum mechanics and thermodynamics arguments. This theory uses the classical balance laws of mass and momenta, together with a new balance law for microforces. This microforce balance takes into account the ``microscopic work'' done by changes in the order parameter, which represents, in the context of our work, the dimensionless concentration of a phase.

The introduction of microforces and microstresses, as well as their respective balance laws, modifies the mechanical version of the second law of thermodynamics [\citet{GUR96b}, Equation (30)], which in turn influences the constitutive relations under consideration. In particular, a byproduct of this framework is that the capillary stress tensor naturally appears as a constraint imposed by the mechanical version of the second law of thermodynamics on the constitutive equations [\citet{GUR96b}, Equation (48)]. \citet{LIU14} extended the framework of Gurtin to encompass the $ {n} $-component Navier--Stokes--Cahn--Hilliard (NSCH) multiphase system with a choice of switching to Van~der~Waals theory.

In this work, we use the dimensionless form of the equations of motion to develop the complete energy budget equation of the coupled Navier--Stokes--Cahn--Hilliard system to model the dynamics of liquid droplets in a liquid continuum. In addition, from a physical point of view, we give an interpretation of each quantity involved in the energy exchange and analyze the behavior of the mass flux across the interface.

From a numerical point of view, we develop a general and robust formulation based on a finite-dimensional, high-order isogeometric analysis approximation. We use divergence-conforming B-spline spaces to obtain a discrete point-wise divergence-free velocity field \citet{EVA13a,EVA13b,EVA13c,SAR15}. We implement this discretization in PetIGA, a high-performance isogeometric analysis framework \citep{DAL15,VIG15b}. Two- and three-dimensional numerical results highlight the robustness of the framework. Finally, we compare our numerical results against analytical features of droplet dynamics to verify our model.

The outline of this work is as follows. Section \ref{sec:gov.eqs} presents the general governing equations, followed by section \ref{sec:con.eqs}, which describes the constitutive relations adopted here. Section \ref{sec:gov.eqs.dimensionless} introduces the dimensionless set of equations together with the dimensionless constitutive equations. Section \ref{sec:mass.flux} details the overall mass flux behavior characterized by the Cahn--Hilliard equation. In section \ref{sec:energy}, we present the energy budget of the Navier--Stokes--Cahn--Hilliard equations under the Bussinesq premise. This concludes the theoretical analysis and preliminaries of the paper. In section \ref{sec:num.scheme} we describe the numerical scheme we use, whereas in section \ref{sec:problem} the numerical experiments setup is detailed. In section \ref{sec:results}, we analyze the numerical results. Finally, in section \ref{sec:conclusions}, we briefly detail the conclusions. \ref{sec:thermodynamics} presents the first and second laws of thermodynamics, showing that our model is thermodynamically consistent. \ref{sec:identities} lists the identities obtained to develop the energy budget of the Navier--Stokes--Cahn--Hilliard flows.

%===============================================================================================================%

\section{General Governing Equations}\label{sec:gov.eqs}

Dynamics of binary immiscible fluids involve both mass and momentum transfer. In this work, the phase segregation phenomenon is described by the Cahn--Hilliard equation \citep{CAH58,CAH59b,CAH59a}, while the hydrodynamics is incorporated through the incompressible Navier--Stokes equations under the Boussinesq assumption to consider buoyancy effects \citep{MEI10,ESP15a}. We assume a small density difference between the fluids. The resulting set of equations in the dimensional form is given by
\begin{subequations}\label{eq:pde.dimensional}
\begin{align}
\divx{\T{v}{1}} = & \, 0, \label{eq:mass} \\
\partn{\phi}{t}{} + \divx{(\phi \T{v}{1})} + \divx{\T{j}{1}} = & \, 0, \label{eq:specie} \\
\rho \left[ \partn{\T{v}{1}}{t}{} + \divx{(\T{v}{1} \otimes \T{v}{1})} \right] - \divx{\T{T}{2}} - \phi \jump{\rho} \T{g}{1} = & \, \T{0}{1}, \label{eq:linear.momentum} \\
\T{T}{2} = & \, \T{T}{2}^T, \label{eq:angular.momentum}
\end{align}
\end{subequations}
where $ {\rho} $, $ {\T{v}{1}} $, $ {\T{T}{2}} $, $ {\phi \jump{\rho} \T{g}{1}} $, $ {\phi} $, and $ {\T{j}{1}} $ are the density, velocity, stress, buoyancy force, volume concentration, and mass flux, respectively. The density is defined as the weighted mean of the fluid densities, $ {\rho = \rho _2 \phi + \rho _1 (1-\phi)} $, where the subscript indicates the corresponding fluid. We only consider the density excess $ {\jump{\rho} = \rho_2 - \rho_1} $ in the buoyancy force. We denote first and second order tensors by bold lower and upper case symbols, respectively. The differential operators $ {\nabla (\cdot)} $ and $ {\nabla\cdot (\cdot)} $ represent the gradient and divergence, respectively. The superscript $ {(\cdot)^T} $ denotes the transpose and $ {\otimes} $ the tensor (dyadic) product.

%===============================================================================================================%

\section{Constitutive Equations}\label{sec:con.eqs}

\subsection*{\centering Mass flux}

According to \citet{LAN59a}, in the absence of heat transfer, the mass flux may be defined by generalizing Fick's law, $ {\T{j}{1} = - \alpha \, \gradx{\eta}} $, where $ {\eta} $ is the chemical potential and $ {\alpha} $ is the chemical mass diffusivity. Once the chemical potential is defined by the Ginzburg--Landau free energy density, i.e., $ {\eta \eqdef {\delta \Psi}/{\delta \phi}} \, [\text{energy}/\text{volume}] $ (the variational derivative of $ {\Psi} $), the Cahn--Hilliard equation is obtained \citep{GOM08,EMM03}. Additionally, the Ginzburg--Landau free energy, in a volume $ {\Omega} $, may be written as
\begin{equation}\label{eq:ginzburg.landau.01}
\Psi[\phi] = \int _{\Omega} \psi \diffo{\Omega} = \int _{\Omega} (\psi_{\phi}+\psi_s) \diffo{\Omega},
\end{equation}
where the bulk and interfacial free energy densities $ {\psi_{\phi}} $ and $ {\psi_s} $, respectively, are defined as [\citet{CAH58}, Equations (3.1) and (3.13)]
\begin{subequations}\label{eq:ginzburg.landau.02}
\begin{align}
\psi_{\phi} &= N_v k_B \theta (\phi \ln \phi + (1-\phi) \ln (1-\phi) ) + N_v \omega (1-\phi)\phi, \\
\psi_s &= \dfrac{\gamma_{\phi}}{2} \, \gradx{\phi} \cdot \gradx{\phi}.
\end{align}
\end{subequations}
Here $ {N_v} $ is the number of molecules per unit volume, $ {k_B} $ is Boltzmann's constant, and $ {\omega} $ is an interaction energy given by $ {\omega = 2k_B \theta_c} $. The interaction energy is positive and is related to the critical temperature, $ {\theta_c} $. In the interfacial free energy density term, $ {\gamma_{\phi} = \sigma l} \, {[\text{force}]} $ represents the magnitude of the interfacial energy. The parameters $ {\sigma} $ and $ {l} $ are the interface tension $ {[\text{force}/\text{length}]} $ and the interfacial thickness\footnote{Alternatively, a root mean square effective ``interaction distance'', according to the original work of \citep{CAH59b}} $ {[\text{length}]} $, respectively. This force term $ {\gamma_{\phi}} $ is defined in [\citet{CAH58}, Equation (3.12)] and [\citet{CAH59b}, Equation (4.1)] by $ {\gamma_{\phi} = N_v \omega l^2} $.

Without loss of generality, we define the interface thickness to have the simplest expression for $ {\gamma_{\phi}} $, i.e., $ {\sigma l} $. \citet{YUE04} opted to define this relation as $ {\gamma_{\phi} = 3 \sigma l /(2\sqrt{2})} $, based on the quartic bulk free energy density when the diffuse interface is at equilibrium ($ {{\delta \Psi}/{\delta \phi} = 0} $).

The scalar parameter $ {\alpha} \, {[\text{time} \! \times \! \text{volume}/\text{mass}]} $ in the mass flux is rewritten as $ {\alpha = M(\phi) \beta} $, where the mobility is $ {M(\phi) = M_o (1-\phi)\phi} $ $ {[\text{length}^2/\text{time}]} $ (with $ {M_o} $ a positive constant) and $ {\beta^{-1}} $ is related to the kinetic energy at a molecular scale, $ {\rho_1 u_m^2 } $ $ {[(\text{mass}/\text{volume}) \! \times \! (\text{length}/\text{time})^{2}]} $. The mobility is degenerate, i.e., the phase dependence of the mobility confines the molecular movement to the interface region. This effect is detailed in Section \ref{sec:mass.flux}. The mass flux can then be defined as
\begin{equation}\label{eq:mass.flux}
\T{j}{1} = - M(\phi) \beta \, \gradx{\eta}.
\end{equation}

Considering the buoyancy effects, we must include the potential energy into the free energy, i.e.,
\begin{equation}\label{eq:ginzburg.landau.03}
\Psi[\phi] = \int _{\Omega} \psi \, \diffo{\Omega} = \int _{\Omega} (\psi_{\phi}+\psi_s+e_p) \diffo{\Omega},
\end{equation}
where $ {e_p= \phi \jump{\rho} g x_2} $ is the potential energy, $ {g} $ is the gravity acceleration, and $ {x_2} $ is the vertical coordinate. Thus, the total mass flux is now defined as
\begin{equation}\label{eq:mass.flux.buoyancy}
\T{j}{1} = - M(\phi) \beta \, \gradx{(\eta_{\phi}+\eta_s)} - M_o \beta \, \gradx{\eta_p},
\end{equation}
where $ {\eta_{\phi}} $, $ {\eta_s} $ and $ {\eta_p} $ are the chemical potentials related to the bulk, interfacial, and potential energies, respectively. The mass flux due to the potential energy is defined with a constant mobility $ {M_o} $, since this is not an interfacial but a bulk mass flux. Thus, for this term the chemical mass diffusivity $ {\alpha} $ is constant.

\subsection*{\centering Stress}

For an incompressible Newtonian fluid, the viscous stress is given by \citep{GUR10}
\begin{equation}\label{eq:constitutive.momentum.01}
\T{T}{2}^{visc} = 2 \mu \T{D}{2}
\end{equation}
with
\begin{equation}\label{eq:strain.rate}
\T{D}{2} = \dfrac{1}{2} \left( (\gradx{\T{v}{1}})^T + \gradx{\T{v}{1}} \right),
\end{equation}
where the dynamic viscosity, $ {\mu \eqdef \mu (\phi)} $ relates the strain rate $ {\T{D}{2}} $ to the viscous stress tensor $ {\T{T}{2}^{visc}} $. The Cauchy stress tensor is defined as $ {\T{T}{2}^{c} = \T{T}{2}^{visc} - p \T{1}{2}} $ \citep{GUR10}, with the last term associating the pressure to the Lagrange multiplier that enforces the incompressibility constraint.

Considering a binary mixture where each phase is endowed with a different viscosity, we assume a smooth transition through the interface which is given by $ {\mu (\phi) = \mu_1 \text{e}^{m \phi}} $ where $ {m = \ln ({\mu_2/\mu_1})} $, being $ {\mu_1} $ and $ {\mu_2} $ the viscosity of each phase. Here, we assume $ {\mu_1 < \mu_2} $. To simplify the exposition in the sequel, we define $ {c = \text{e}^{m \phi}} $.

Using the balance of microforces, \citet{GUR96b} derived a thermodynamically consistent complement to the constitutive relation for the Cauchy stress via the mechanical version of the second law of thermodynamics, which models capillarity effects. The interfacial capillary stress assumes the following form
\begin{equation}\label{eq:constitutive.momentum.02}
\T{T}{2}^{s} = - \gamma_{\phi} \, \gradx{\phi} \otimes \gradx{\phi}.
\end{equation}
Finally, the total stress associated to the macroscopic motion of the fluid is given by
\begin{equation}\label{eq:stress}
\T{T}{2} = \T{T}{2}^{c} + \T{T}{2}^{s} = 2 \mu \T{D}{2} - p \T{1}{2} - \gamma_{\phi} \, \gradx{\phi} \otimes \gradx{\phi},
\end{equation}
where the stress has units of $ {[\text{force}/\text{length}^2]} $.

Finally, our model differs from Gurtin's model in the sense that we account for the potential energy in the free energy, and we assume that the density difference is small. This assumption justifies the use of the Boussinesq assumption in the momentum equation to account for buoyancy effects. In addition, we employ a logarithmic function to represent the free energy rather than the more often found polynomial approximation. In this sense, our work also differs from those presented by \cite{JAC99} and \cite{JAM01}, where the double-well polynomial function is used to describe the free energy, and where the potential energy is not included in the free energy description.

%===============================================================================================================%

\section{Governing Equations in Dimensionless Form}\label{sec:gov.eqs.dimensionless}

To make the governing equations dimensionless, a length $ {b} $ and a velocity $ {u} $ are chosen as the characteristic length and velocity scales, respectively. The reference length and velocity are usually chosen as the droplet diameter or radius and the terminal velocity or buoyancy velocity, respectively. We scale the viscosities by the smallest viscosity $ {\mu_1} $, the pressure by $ {\mu_1 u/b} $, the time by $ {b/u} $, while the mobility is scaled by $ {M_o} $. The chemical potential is rendered dimensionless by $ {\psi_c = 2N_v k_B \theta_c = N_v \omega} $, which is the critical free energy density. The phase field is inherently a dimensionless quantity and is normalized between $ {(0,1)} $ according to the bulk free energy density. The mixture law assumes the density as $ {\rho = \rho_1 + \phi \jump{\rho}} $, being $ {\jump{\rho} = \rho_2 - \rho_1} $. In our examples, $ \rho_1 $ is the ligther fluid, nevertheless, the association between lower viscosity and lower density is incidental and can be reversed if the modeling requires this. In addition, due to the Boussinesq assumption we have that $ {\mathcal{O} (\frac{\phi \jump{\rho}}{\rho_1}) \ll 1} $, usually in the order of $ {5\%} $ would be admissible. Table \ref{tb:dimensionless.number} lists the dimensionless numbers we obtain with these scalings.
\begin{table*}[!t]
  \caption{Dimensionless Groups}
  \label{tb:dimensionless.number}
  \centering
  \footnotesize
  \begin{tabular}{l l l l}
  \toprule
  Number   &   & Equation                          & Interpretation \\ %[0.5cm]
  \midrule
  Peclet   & - & $ Pe = ub/M_o $                   & inertia/mass diffusion \\ %[0.5cm]
  Reynolds & - & $ Re = \rho_1 ub/\mu_1 $          & inertia/momentum diffusion \\ %[0.5cm]
  Weber    & - & $ We = \rho_1 u^2b/\sigma $       & inertia/surface tension \\ %[0.5cm]
  Bond     & - & $ Bo = \jump{\rho} g b^2/\sigma $ & potential energy/interfacial energy \\ %[0.5cm]
  Cahn     & - & $ Cn = l/b $                      & interfacial thickness/length scale \\ %[0.5cm]
  $ Lm $   & - & $ Lm = \psi_c b/\sigma $          & critical free energy density/capillary energy \\ %[0.5cm]
  $ Ln $   & - & $ Ln = \rho_1 u_m^2/\psi_c $      & kinetic energy at the molecular scale/critical free energy density \\ %[0.5cm]
  \bottomrule
  \end{tabular}
\end{table*}

To the best of our knowledge, two new dimensionless groups appear in this physical interpretation. We denote these new numbers as $ {Lm} $ and $ {Ln} $. $ {Lm} $ is the ratio between the critical free energy density and the capillary energy, i.e., the interfacial curvature energy. The capillary energy is closely related to the Laplace pressure obtained from the Young--Laplace equation \citep{MYE90}. This pressure difference (or Laplace pressure) $ {\jump{p} = \sigma ( R_{1}^{-1} + R_{2}^{-1})} $ relates the surface tension with the principal radii of curvature. In the particular case of a spherical droplet $ {\jump{p} = 2 \sigma R^{-1}} $, where $ {R} $ is the radius of the sphere. The interfacial curvature energy takes the form $ {\sigma/b = \jump{p} \! / 4} $. $ {Ln} $ is the ratio between the kinetic energy at a molecular scale and the critical free energy density. In the particular case of an ideal gas, $ {Ln} $ can be interpreted by defining the molecular velocity $ {u_m} $ as $ {u_m^{2} = u_{rms}^{2} = 3k_B \theta/m} $ where $ {u_{rms}} $ is the root-mean-square speed of a single molecule of mass $ {m} $. This assumption yields $ {Ln = \rho_1 u_{rms}^2/\psi_c = 3/2 \vartheta^{-1}} $, where $ {\vartheta = \theta_c / \theta} $ defines the ratio between the critical and the absolute temperatures. Therefore, $ {Ln} $ can measure the deviation of the absolute temperature from the critical one in an ideal gas.

Henceforth, all quantities considered are dimensionless according to Table \ref{tb:dimensionless.number}. The dimensionless forms of the free energy densities are
\begin{subequations}\label{eq:dimensionless.free.energies}
\begin{align}
\psi_{\phi} = & \dfrac{1}{2\vartheta} (\phi \ln \phi + (1-\phi) \ln (1-\phi) ) + (1-\phi)\phi, \\
\psi_s = & \dfrac{Cn}{2Lm} \gradx{\phi} \cdot \gradx{\phi}, \\
e_p = & \dfrac{Bo}{We} \, \phi x_2,
\end{align}
\end{subequations}
and the chemical potential reads
\begin{equation}\label{eq:dimensionless.chemical.potential}
\eta = \dfrac{\delta \Psi}{\delta \phi} = \partn{\psi}{\phi}{} - \divx{\partn{\psi}{\gradx{\phi}}{}} = \dfrac{1}{2\vartheta} \ln \dfrac{\phi}{1-\phi} + 1 - 2\phi - \dfrac{Cn}{Lm} \, \Delta\phi + \dfrac{Bo}{We} \, x_2.
\end{equation}
The set of dimensionless differential equations is given by
\begin{subequations}\label{eq:dimensionless.pde}
\begin{align}
\divx{\T{v}{1}} = & \, 0, \label{eq:dimensionless.mass} \\
\partn{\phi}{t}{} + \T{v}{1} \cdot \gradx{\phi} + \divx{\T{j}{1}} = & \, 0, \label{eq:dimensionless.specie} \\
\partn{\T{v}{1}}{t}{} + \T{v}{1} \cdot \gradx{\T{v}{1}} - \divx{\T{T}{2}} + \dfrac{Bo}{We} \phi \, \T{e}{1}_2  = & \, \T{0}{1}, \label{eq:dimensionless.momentum}
\end{align}
\end{subequations}
where the dimensionless mass flux is defined as
\begin{equation}\label{eq:dimensionless.mass.flux}
\T{j}{1} = - \dfrac{1}{PeLn} \, \left\{ M(\phi) \, \gradx{\left(\dfrac{1}{2\vartheta} \ln \dfrac{\phi}{1-\phi} + 1 - 2\phi - \dfrac{Cn}{Lm} \, \Delta\phi \right)} + \dfrac{Bo}{We} \, \T{e}{1}_2 \right\},
\end{equation}
the stress is given by
\begin{equation}\label{eq:dimensionless.stress}
\T{T}{2} = \dfrac{2c}{Re}\T{D}{2} - \dfrac{1}{Re} p \T{1}{2} - \dfrac{Cn}{We} \, \gradx{\phi} \otimes \gradx{\phi},
\end{equation}
and $ {c} $ scales the kinetic viscosities of the different fluids. The rescaled mobility is equal to $ {M(\phi) = \phi(1-\phi)} $. For convenience, we assume gravity points in the -$ {\T{e}{1}_2} $ direction.

To understand the components of the mass flux, we split the chemical potential into $ {\eta = \eta_{\phi} + \eta_s + \eta_p} $ where $ {\eta_{\phi} = \frac{1}{2\vartheta} \ln \frac{\phi}{1-\phi} + 1 - 2\phi} $, $ {\eta_s = - \frac{Cn}{Lm} \, \Delta\phi} $ and $ {\eta_p = \frac{Bo}{We} \, x_2} $, which yield the bulk, interfacial, and gravitational mass fluxes, respectively.

%===============================================================================================================%

\section{Mass Flux Behavior}\label{sec:mass.flux}

In the Cahn--Hilliard equation, the mass flux is a phobic phenomenon that segregates the phases. The mass flux exhibits an ``anomalous'' behavior across the interface thickness, as it may change its sign (twice) and deviate from the gradient direction of the phase field. In the following, we present the analysis of the orientation of these mass fluxes. We start by analyzing the flux structure in a one-dimensional example. Next, we describe some of the restrictive conditions under which both mass flux terms are parallel. We conclude this section with a detailed description of a general two-dimensional simulation. To later explain the energy exchanges in the free energy densities, we first develop a good understanding of the mass flux behavior. To do so, we rewrite the mass flux (\ref{eq:dimensionless.mass.flux}) as
\begin{equation}\label{eq:anomalous.mass.flux}
\begin{split}
\T{j}{1} = & -\dfrac{1}{PeLn} \, M(\phi) \, \gradx{(\eta_{\phi}+\eta_s)} -\dfrac{1}{PeLn} \dfrac{Bo}{We} \, \gradx{\eta_p} \\
= & - \dfrac{M(\phi)}{PeLn} \left[ \left( - 2 + \dfrac{1}{2\vartheta \phi (1-\phi)} \right) \gradx{\phi} - \dfrac{Cn}{Lm} \, \divx{\T{H}{2}} \right] - \dfrac{1}{PeLn} \dfrac{Bo}{We} \, \T{e}{1}_2. \\
\end{split}
\end{equation}
To simplify notation, we define the bulk, $ {\T{j}{1}_{\phi}} $, interfacial, $ {\T{j}{1}_{s}} $, and gravitational, $ {\T{j}{1}_{p}} $, mass fluxes as
\begin{subequations}\label{eq:mass.fluxes}
\begin{align}
\T{j}{1}_{\phi} = & - \dfrac{1}{PeLn} \left( - 2 + \dfrac{1}{2\vartheta \phi (1-\phi)} \right) M(\phi) \, \gradx{\phi}, \\
\T{j}{1}_{s} = & \dfrac{1}{PeLn} \dfrac{Cn}{Lm} \, M(\phi) \, \divx{\T{H}{2}}, \\
\T{j}{1}_{p} = & - \dfrac{1}{PeLn} \dfrac{Bo}{We} \, \T{e}{1}_2,
\end{align}
\end{subequations}
where $ {\T{H}{2}} $ denotes the Hessian of $ {\phi} $. In general, $ {\gradx{\eta_{\phi}}} $ and $ {\gradx{\eta_s}} $ are not parallel, thus the mass flux might not be normal to the interface implicitly defined by isovalues of the phase field. 

\begin{figure}[!t]
\centering
  \includegraphics[width=1.0\textwidth]{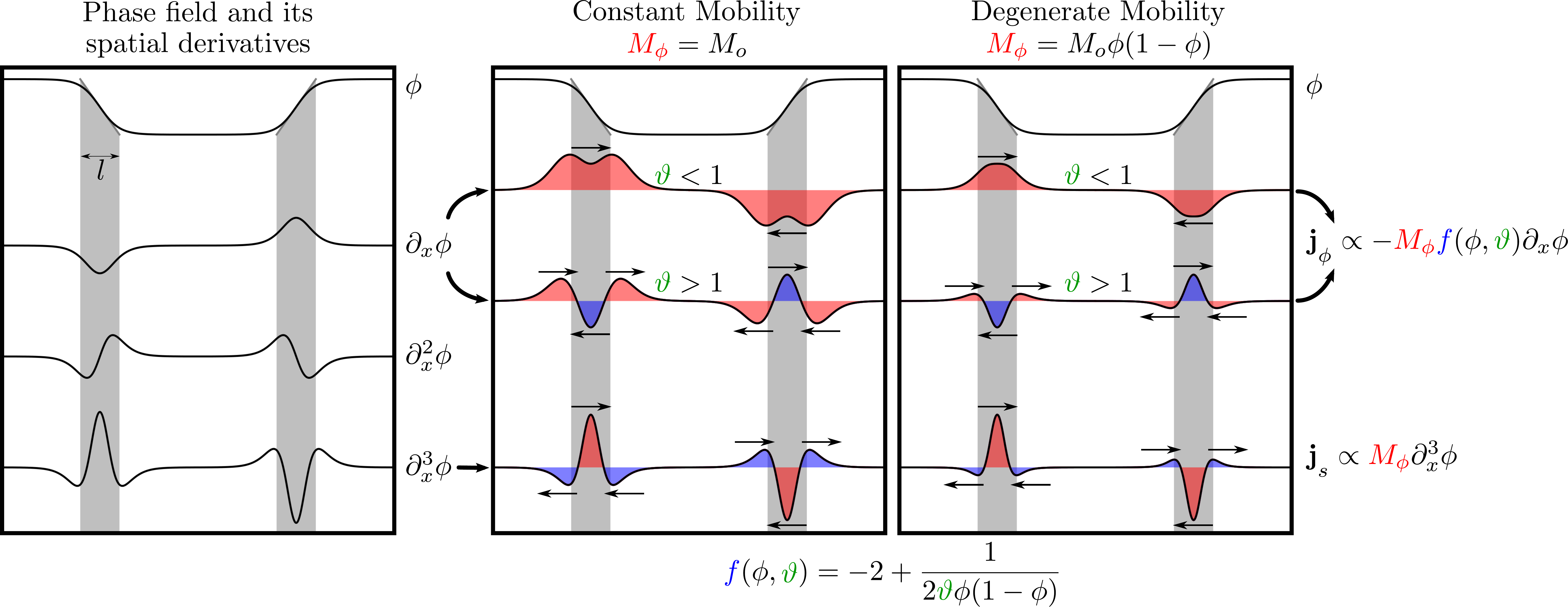}
  \caption{(color online) Phase field and its spatial derivatives in one spatial dimension in the first diagram (left). Mass fluxes with constant mobility in the second diagram (middle) are depicted considering both cases, $ {\vartheta < 1} $ and $ {\vartheta > 1} $ ($ {\vartheta = \theta _c/ \theta} $) whereas the same mass fluxes are depicted with degenerate mobility in the third diagram (right).}
\label{fg:phase.and.derivatives}
\end{figure}

First, let us observe how the mass flux may change sign across the interface. Figure \ref{fg:phase.and.derivatives} shows, in the first column, the phase field and its derivatives up to third order in a one-dimensional problem. In this Figure, lower values of $ {\phi} $ represent the droplet (inclusion). Inward and outward refer with respect to this region. In addition, Figure \ref{fg:phase.and.derivatives} shows the mass flux behavior assuming constant and degenerate mobilities (second and third columns, respectively). In all columns the first row shows the phase field. The second and third rows in the second and third columns show the dependence of the bulk mass flux $ {\T{j}{1}_{\phi}} $ on the temperature ratio $ {\vartheta} $. Finally, the fourth row depicts the interfacial mass flux $ {\T{j}{1}_s} $. Red indicates the regions where the mass flux points inward, whereas blue indicates the regions where the mass flux points outward. The bulk mass flux, $ {\T{j}{1}_{\phi}} $, points in the steepest descent direction if the temperature ratio is lower than unity, $ {\vartheta < 1} $, while this flux changes its sign twice if the temperature ratio is greater than one, $ {\vartheta > 1} $. On the other hand, the interfacial mass flux $ {\T{j}{1}_s} $ always exhibits two changes of sign. Thus, the total mass flux $ {\T{j}{1}} $ may have two changes in sign. These changes of the mass flux direction explain the phobic interactions between the phases. The function $ {- 2 + {1}/{(2\vartheta \phi (1-\phi))}} $ is positive definite if $ {\vartheta < 1} $, otherwise if $ {\vartheta > 1} $ it becomes indefinite, being $ {\vartheta} $ a finite value. Finally, if $ {\vartheta \rightarrow \infty} $ the function is negative definite.

\begin{rmk}[The overall behavior of the mass fluxes]\label{rmk:mass.flux.0}
The overall behavior of the $ {\divx{\T{H}{2}}} $ is such that it has a positive inner product with $ {-\gradx{\phi}} $ when Neumann-free boundary conditions are considered. For $ {\gradx{\phi} \cdot \T{n}{1} = 0 \vert_{\Gamma}} $, being $ {\T{n}{1}} $ the outward unit normal vector to the boundary $ {\Gamma} $, we obtain $ {\int_{\Omega} - \gradx{\phi} \cdot (\divx{\T{H}{2}}) \, \diffo{\Omega} = \int_{\Omega} (\Delta\phi)^2 \, \diffo{\Omega}} $. \qed
\end{rmk}

\begin{rmk}[Mass flux in two dimensions]\label{rmk:mass.flux.1}
We consider an idealized two-dimensional case to understand how the interfacial mass flux \emph{$ {\T{j}{1}_s} $} may deviate the total mass flux from the steepest descent direction of the phase field. Let us define a local coordinate system given by the Frenet basis \emph{$ {(\T{e}{1}_{\perp},\T{e}{1}_{\parallel})} $} \citep{KUH06}. The basis is defined over isovalue curves of \emph{$ {\phi} $}, being \emph{$ {\T{e}{1}_{\perp}} $} the unit vector in the direction of the gradient of \emph{$ {\phi} $}, \emph{$ {n} $}, and \emph{$ {\T{e}{1}_{\parallel}} $} the unit vector in the tangential direction, \emph{$ {\tau} $}. We describe the coordinate transformation between Frenet and Cartesian basis by taking the vector \emph{$ {\diffo{\T{x}{1}} = (\diffo{x}_1,\diffo{x}_2)} $} and establishing that
\emph{
\begin{equation}\label{eq:jacobian.basis}
\diffo{\T{x}{1}} = \spart{\T{x}{1}}{n}{} \diffo{n} + \spart{\T{x}{1}}{\tau}{} \diffo{\tau},
\end{equation}
}
where \emph{$ {\spart{}{n}{} = \frac{\partial}{\partial n}} $} and \emph{$ {\spart{}{\tau}{} = \frac{\partial}{\partial \tau}} $}. We set the metric coefficients as the normal \emph{$ {h_n \eqdef \vert \spart{\T{x}{1}}{n}{} \vert} $} and tangential \emph{$ {h_{\tau} \eqdef \vert \spart{\T{x}{1}}{\tau}{} \vert} $} components. Thus, the gradient operator, the gradient of \emph{$ {\phi} $}, the Laplacian operator, and the Laplacian of \emph{$ {\phi} $} are defined in this coordinate system as
\emph{
\begin{subequations}\label{eq:gradient.frenet}
\begin{align}
\gradx{(\cdot)} = & \, h^{-1}_n \spart{(\cdot)}{n}{} \, \T{e}{1}_{\perp} + h^{-1}_{\tau} \spart{(\cdot)}{\tau}{} \, \T{e}{1}_{\parallel}, \\
\gradx{\phi} = & \, h^{-1}_n \spart{\phi}{n}{} \, \T{e}{1}_{\perp}, \\
\Delta(\cdot) = & \, (h_nh_{\tau})^{-1} \left[ \spart{ (h_{\tau}h^{-1}_n \spart{(\cdot)}{n}{})}{n}{} + \spart{ (h^{-1}_{\tau}h_n \spart{(\cdot)}{\tau}{})}{\tau}{} \right], \\
\Delta\phi = & \, (h_nh_{\tau})^{-1} \spart{ (h_{\tau}h^{-1}_n \spart{\phi}{n}{})}{n}{}.
\end{align}
\end{subequations}
}
Finally, the divergence of the Hessian of \emph{$ {\phi} $} in this orthogonal curvilinear coordinate system is
\emph{
\begin{equation}\label{eq:div.hessian.frenet.01}
\begin{split}
\divx{\T{H}{2}} = & \, h^{-1}_n \left\{ \spart{(h_nh_{\tau})^{-1}}{n}{} [\spart{(h^{-1}_{n}h_{\tau} \spart{\phi}{n}{})}{n}{}] + (h_nh_{\tau})^{-1} [\spart{(h^{-1}_{n}h_{\tau} \spart{\phi}{n}{})}{n}{2}] \right\} \, \T{e}{1}_{\perp} \\
+ & \, h^{-1}_{\tau} \left\{ \spart{(h_nh_{\tau})^{-1}}{\tau}{} [\spart{(h^{-1}_{n}h_{\tau} \spart{\phi}{n}{})}{n}{}] + (h_nh_{\tau})^{-1} [\spart{\spart{(h^{-1}_{n}h_{\tau} \spart{\phi}{n}{})}{n}{}}{\tau}{}] \right\} \T{e}{1}_{\parallel}.
\end{split}
\end{equation}
}
By assuming that isocurves of \emph{$ {\phi} $} are parallel (i.e., constant interface thickness) and the tangential metric along \emph{$ {\tau} $} is constant (i.e., the isocurves of \emph{$ {\phi} $} are circumferences or straight lines), we conclude that (\emph{$ {\spart{\T{e}{1}_{\perp}}{n}{} = \spart{\T{e}{1}_{\parallel}}{n}{} = \T{0}{1}} $ $ {\Rightarrow} $ $ {\spart{h_n}{n}{} = \spart{h_{n}}{\tau}{} = 0} $}) and (\emph{$ {\spart{h_{\tau}}{\tau}{} = 0} $}), respectively. Thus, under these premises, we write the divergence of the Hessian of \emph{$ {\phi} $} as
\emph{
\begin{equation}\label{eq:div.hessian.frenet.02}
\divx{\T{H}{2}} = \varphi \, \T{e}{1}_{\perp}
\end{equation}
}
being \emph{$ {\varphi = h^{-1}_n \spart{h_{\tau}}{n}{} \left[ \spart{h_{\tau}}{n}{} \spart{\phi}{n}{} + h_{\tau} \spart{\phi}{n}{2} + (h^2_nh_{\tau})^{-1}\spart{h_{\tau}}{n}{2} \spart{\phi}{n}{} + h_{n}^{-2}\spart{\phi}{n}{3} \right]} $}. \qed
\end{rmk}

The idealized example explained above deals with isocontours of $ \phi $ endowed with parallel and constant curvature to show the simplest behavior of the mass flux, yielding a mass flux normal to the interfaces. In the general case, the mass flux deviates from the normal direction due to changes in the metric coefficients in both directions.

\begin{rmk}[Mass flux in steady state solutions]\label{rmk:steady.cahn.hilliard}
From Remark \ref{rmk:mass.flux.1}, we conclude that the steady state solution of the Cahn--Hiliard equation yields \emph{$ {\gradx{\phi} \parallel \divx{\T{H}{2}}} $}. We stress that this conclusion does not consider hydrodynamic effects.
\end{rmk}

\begin{figure}[!t]
\centering
  \includegraphics[width=0.6\textwidth]{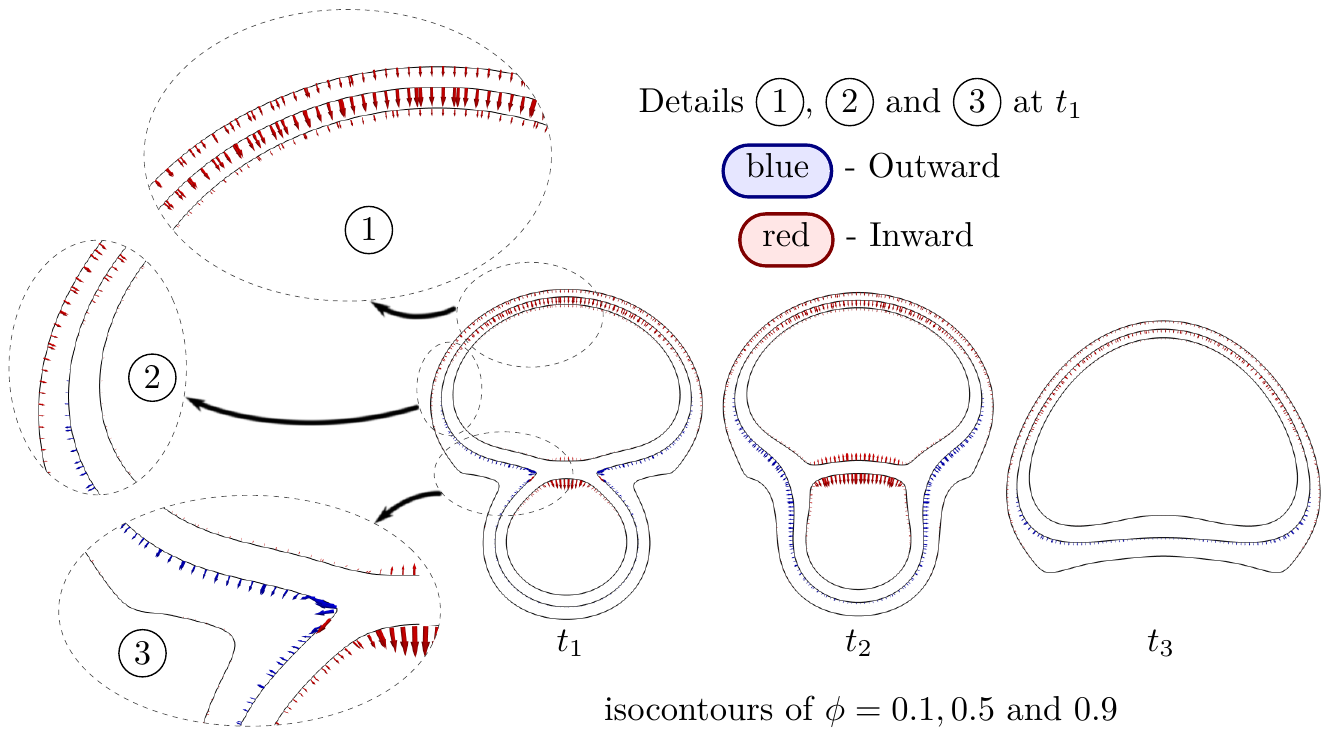}
\caption{(color online) Mass flux over isocontours of $ {\phi = 0.1, 0.5} $ and $ {0.9} $ at three instants $ {t_1} $, $ {t_2} $ and $ {t_3} $ when two droplets are merging and rising. Three details at $ {t_1} $ present regions where the mass flux changes its sign and deviates from the normal direction.}
\label{fg:mass.flux}
% \end{figure}
\vspace{0.5cm}
% \begin{figure}
\centering
  \includegraphics[width=0.6\textwidth]{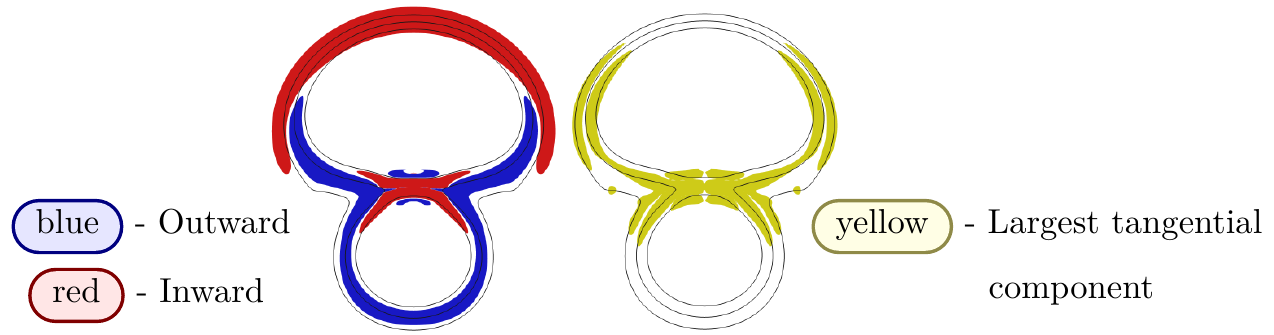}
\caption{(color online) On the left, red (blue) depicts the region where the mass flux points inward (outward) and on the right, yellow depicts the region where the tangential component of the mass flux is largest.}
\label{fg:norm.tang}
\end{figure}

Figure \ref{fg:mass.flux} shows three stages in the merging of two rising droplets and how the mass flux behaves during this process. At an early stage, details $ {1} $ through $ {3} $ depict the mass flux on three isocontours of the phase field, $ {\phi = 0.1, 0.5} $ and $ {0.9} $. Across the interface, in the leading zone of the top droplet, the mass flux points inward and is almost normal to the interface, cf. detail $ {1} $. However, detail $ {2} $ shows a gradual change in the mass flux direction. In this region, the mass flux points outward on a isocontour $ {\phi = 0.5} $, whereas on a isocontour $ {\phi = 0.9} $ the mass flux points inward. In the regions where the phase field isocontours exhibit a large change in curvature, which happens when droplets merge, we observe a significant deviation of the mass flux from the steepest descent direction, as detail $ {3} $ depicts. In the trailing zone, the mass flux points outward during the entire merging process. The largest values of the tangential component of the mass flux are located in the highly curved region of the phase field. At the last stage, instant $ {t_3} $, large mass flux deviations are not observed.

On the left of Figure \ref{fg:norm.tang}, the red (blue) region represents where the mass flux points inward (outward) with respect to the droplets. In the leading (trailing) zone, the mass flux points inward (outward). In the highly curved region, the mass flux pushes the droplet shape to reduce its curvature. In the contact region between the droplets, the mass flux points toward the interiors of the droplets and the mass flux changes its sign across the contact interface between two droplets. On the right of Figure \ref{fg:norm.tang}, the regions where the tangential component is large are shown in yellow. Figures \ref{fg:mass.flux} and \ref{fg:norm.tang} were obtained from the second example (case $ {\#2} $) detailed in Section \ref{sec:results}. This example involves the merging of two rising liquid droplets embedded in a continuous liquid. Finally, as $ {\T{j}{1}_p} $ is a divergence-free field, it does not play any role in the mass transfer equation. However, that is not the case for the energy budget equation, as we will show in the following section.

%===============================================================================================================%

\section{Energy Budget}\label{sec:energy}

In this section, we describe the temporal evolution of the different energy components of the flow and analyze their physical meaning. The energy budget plays a central role in the scientific understanding of physical phenomena. We analyze the exchanges of the kinetic, potential, bulk, and interfacial free energies. In particular, the energy budget describes the energy exchange mechanisms that govern the flow. Moreover, according to \citet{LIU00,GOU14}, the convergence of finite-dimensional approximations to the solutions of the partial differential equations is also linked to those energy transfers, particularly for non-smooth solutions.

For the sake of clarity, subscripts $ {k} $, $ {\phi} $, $ {s} $, and $ {p} $ refer in the sequel to quantities derived from the kinetic energy, bulk free energy density $ {\psi_{\phi}} $, interfacial free energy density $ {\psi_s} $, and potential energy, respectively. Assuming that the control volume $ {\Omega} $ does not deform, in the following subsections we derive energy exchanges in an Eulerian description. The identities and definitions used here are detailed in \ref{sec:identities}.

\subsection*{\centering Energies}

The different energies relevant to the incompressible Navier--Stokes--Cahn--Hilliard equations under the Bussinesq assumption in the dimensionless form are
\begin{subequations}\label{eq:energies}
\begin{align}
E_k = & \int _{\Omega} e_k \, \diffo{\Omega} = \int _{\Omega} \dfrac{1}{2} \T{v}{1} \cdot \T{v}{1} \, \diffo{\Omega}, \label{eq:energy.k} \\
E_{\phi} = & \int _{\Omega} \psi_{\phi} \, \diffo{\Omega} =  \int _{\Omega} \left( \dfrac{1}{2\vartheta} (\phi \ln \phi + (1-\phi) \ln (1-\phi) ) + (1-\phi)\phi \right) \, \diffo{\Omega}, \label{eq:energy.phi} \\
E_s = & \int _{\Omega} \psi_s \, \diffo{\Omega} =  \int _{\Omega} \left( \dfrac{Cn}{2Lm} \, \gradx{\phi} \cdot \gradx{\phi} \right) \, \diffo{\Omega}, \label{eq:energy.s} \\
E_p = & \int _{\Omega} e_p \, \diffo{\Omega} = \int _{\Omega} \dfrac{Bo}{We} \, \phi x_2 \, \diffo{\Omega}. \label{eq:energy.p}
\end{align}
\end{subequations}

Consider the material derivative $ {\matd{(\cdot)}} $ of each energy involved in the problem, i.e., $ {\matd{e}_{\jmath}} $ where $ {\jmath = \{k,\phi,s,p\}} $. Integrating these material derivatives over the domain results in the total derivative $ {\frac{\text{d}(\cdot)}{\text{d}t}} $ for each energy in (\ref{eq:energies}). In the following, we consider the equations formulated for a given, undeformed control volume $ {\Omega} $ and boundary $ {\Gamma} $, as
\begin{equation}\label{eq:general.budget}
\diffn{E_{\jmath}}{t}{} = S_{\jmath} + Q_{\jmath} = \int _{\Gamma} \epsilon_{\jmath}^{io} \, \diffo{\Gamma} + \int _{\Omega} \epsilon_{\jmath} \, \diffo{\Omega},
\end{equation}
where, as before, $ {\jmath = \{k,\phi,s,p\}} $ denotes the energy type. The term $ {Q_{\jmath}} $ incorporates the sources and sinks of the energy $ {E_{\jmath}} $ in the control volume, and $ {S_{\jmath}} $ is the total diffusive flux of energy across the boundary of the control volume. Integrating over time the total derivative of each energy, i.e., $ {S_{\jmath} + Q_{\jmath}} $, we obtain the quantities that describe energy exchanges. We denote the extensive (intensive) quantities by uppercase (lowercase) letters, such as, $ {E_{(\cdot)}} $ ($ {e_{(\cdot)}} $) for the extensive (intensive) energies and $ {\mathcal{E}_{(\cdot)}} $ ($ {\epsilon_{(\cdot)}} $) for the extensive (intensive) energy exchanges, i.e., extensive (intensive) sources/sinks of energy.

\subsection*{\centering Kinetic Energy Exchange}

The form of the flux and source/sink terms in the case of the kinetic energy stemming from (\ref{eq:general.budget}) can be derived using the momentum equations and are given by
\begin{subequations}\label{eq:kinetic.dissipation}
\begin{align}
Q_k = & - \int _{\Omega} \T{D}{2} : \left[ \T{T}{2}^{visc} + \T{T}{2}^{s} \right] \, \diffo{\Omega} - \int _{\Omega} \dfrac{Bo}{We} \phi \, v_2 \, \diffo{\Omega}, \nonumber \\
= & - \int _{\Omega} \T{D}{2} : \left[ \dfrac{2c}{Re} \T{D}{2} - \dfrac{Cn}{We} \, \gradx{\phi} \otimes \gradx{\phi} \right] \diffo{\Omega} - \int _{\Omega} \dfrac{Bo}{We} \phi \, v_2 \, \diffo{\Omega}, \\
S_k =  & \int _{\Gamma} \left( \T{v}{1} \cdot \T{T}{2} \right) \cdot \T{n}{1} \, \diffo{\Gamma}, \nonumber \\
= & \int _{\Gamma} \left[ \T{v}{1} \cdot \left( \dfrac{2c}{Re} \T{D}{2} - \dfrac{p}{Re} \T{1}{2} - \dfrac{Cn}{We} \, \gradx{\phi} \otimes \gradx{\phi} \right) \right] \cdot \T{n}{1} \, \diffo{\Gamma}.
\end{align}
\end{subequations}

\emph{Physical interpretation}: Regarding the volumetric term $ {Q_k} $, the first term, $ {\T{D}{2} : \T{T}{2}^{visc}} $, is the rate of work done by the viscous forces of the flow. Given the constitutive model assumed for the viscous stress, Equation (\ref{eq:dimensionless.stress}), the strain rate and the viscous stress have the same eigenvectors, thus the inner product is always positive, causing the system to lose kinetic energy from the friction between fluid particles. Thus, the first term $ {\T{D}{2} : \T{T}{2}^{visc}} $ is a sink, as can be inferred from its negative sign.

The second term $ {\T{D}{2} : \T{T}{2}^{s}} $ is the rate of work done by the capillary forces in the diffuse interface and can be thought of as follows. The vector $ {\gradx{\phi}} $ is parallel to the normal $ {n} $ of the isosurfaces of $ {\phi} $. We define a local coordinate system at a point on one of these surfaces. In this local coordinate system, the capillary tensor is simply $ {( \frac{\partial \phi}{\partial n} )^2} $, whereas the strain rate contribution is $ {\frac{\partial v_n}{\partial n}} $, with $ {v_n} $ the normal velocity. The contribution to the change of kinetic energy density around this point is then $ {\frac{\partial v_n}{\partial n} ( \frac{\partial \phi}{\partial n} )^2} $.

Suppose $ {v_n} $ grows in the direction of the normal, which by definition is pointing towards the gradient of the phase field. This means that the interfacial thickness will grow, leading to more fluid particles diffusing and becoming a source of kinetic energy by their movement. On the contrary, if $ {v_n} $ grows in the direction opposite to the normal, then this will cause the interface to narrow, thus stripping fluid particles in the diffuse interface of their motion and leading to a loss of kinetic energy. The energy exchange done by the capillary tensor behaves as both a source or a sink of energy, since eigenvectors of the strain rate and the eigenvector of the capillary stress associated with the non-zero eigenvalue (since the capillary stress tensor is rank-one) can have either a negative or positive inner product. In addition, due to incompressibility, the pressure does not play any role in $ {Q_k} $. Since we use divergence-conforming discretizations, the orthogonality of the pressure and the strain rate is preserved at the discrete level in our numerical simulations.

The first and second terms in the kinetic energy exchange can also be rewritten as the power expenditures carried out in an infinitesimal increment of the strain rate $ {\diffo{\T{D}{2}}} $ done by the stresses $ {(\T{T}{2}^{visc} + \T{T}{2}^{s}) : \diffo{\T{D}{2}}} $, i.e., $ {Q_k} $ is the rate of work done by the viscous and capillary stresses to achieve an increment $ {\diffo{\T{D}{2}}} $ in strain rate. We give a detailed description of the third term when we interpret the potential energy exchanges below.

Moreover, the boundary term $ {S_k} $ represents the rate of work on the boundary given by the external power done by the total stress.

\subsection*{\centering Bulk Free Energy Exchange}

The form of the terms in (\ref{eq:general.budget}) for the bulk free energy is
\begin{subequations}\label{eq:bulk.dissipation}
\begin{align}
Q_{\phi} = & \int _{\Omega} \gradx{\eta_{\phi}} \cdot \T{j}{1} \, \diffo{\Omega}, \nonumber \\
= & - \int _{\Omega} \dfrac{1}{PeLn} \gradx{\eta_{\phi}} \cdot \left[ M(\phi) \, \gradx{(\eta_{\phi}+\eta_s)} + \gradx{\eta_p} \right] \, \diffo{\Omega}, \\
S_{\phi} = & - \int _{\Gamma} \eta_{\phi} \, \T{j}{1} \cdot \T{n}{1} \, \diffo{\Gamma}, \nonumber \\
= & \int _{\Gamma} \dfrac{1}{PeLn} \eta_{\phi} \left[ M(\phi) \, \gradx{(\eta_{\phi} + \eta_s)} + \gradx{\eta_p} \right] \cdot \T{n}{1} \, \diffo{\Gamma}.
\end{align}
\end{subequations}

\emph{Physical interpretation}: In the volumetric term $ {Q_{\phi}} $, the power expenditure carried out in an infinitesimal increment of the bulk free energy gradient $ {\diffo{\, \gradx{\eta_{\phi}}}} $ done by the mass flux is given by $ {\T{j}{1} \cdot \diffo{\, \gradx{\eta_{\phi}}}} $. Given an infinitesimal increment $ {\diffo{\, \gradx{\eta_{\phi}}}} $, the local exchange of bulk free energy is proportional to the mass flux in the direction of that increment. That is, $ {Q_{\phi}} $ is the rate of work done by the mass flux to achieve an increment in bulk free energy gradient. This term may act as either a source or a sink of energy.

The boundary term $ {S_{\phi}} $ represents the bulk free energy diffusion on the boundaries by the mass flux.

\begin{rmk}\label{rmk:bulk.dissipation}
Replacing the mass flux Equation (\ref{eq:mass.flux}) in Equation (\ref{eq:bulk.dissipation}) and using the Gauss divergence theorem, we find where the sources and sinks of bulk free energy come from. The volumetric $ {Q_{\phi}} $ and boundary $ {S_{\phi}} $ terms take the following form in terms of $ {\phi} $:
\emph{
\begin{subequations}\label{eq:bulk.dissipation.alternative}
\begin{align}
Q_{\phi} = & - \int _{\Omega} \dfrac{1}{PeLn} \Biggl\{ \phi (1-\phi) \left( -2 + \dfrac{1}{2 \vartheta \phi (1-\phi)} \right)^2 \gradx{\phi} \cdot \gradx{\phi} \nonumber \\
+ & \dfrac{Cn}{Lm} \left[ \left( -2\phi (1-\phi) + \dfrac{1}{2 \vartheta} \right) \T{H}{2} : \T{H}{2} - 2 (1 - 2 \phi) \T{H}{2} : \gradx{\phi} \otimes \gradx{\phi} \right] \nonumber \\
+ & \dfrac{Bo}{We} \left( -2 + \dfrac{1}{2 \vartheta \phi (1-\phi)} \right) \gradx{\phi} \cdot \T{e}{1}_2 \Biggr\} \, \diffo{\Omega}, \\
S_{\phi} = & \int _{\Gamma} \dfrac{1}{PeLn} \Biggl\{ \left( \dfrac{1}{2\vartheta} \ln \dfrac{\phi}{1-\phi} + 1 - 2\phi \right) \biggl[ \left( - 2 \phi (1-\phi) + \dfrac{1}{2\vartheta} \right) \gradx{\phi} \nonumber \\
- & \dfrac{Cn}{Lm} \phi (1-\phi) \, \divx{\T{H}{2}}  + \dfrac{Bo}{We} \T{e}{1}_2 \biggr] + \dfrac{Cn}{Lm} \left( -2 \phi (1-\phi) + \dfrac{1}{2 \vartheta} \right) \gradx{\phi} \cdot \T{H}{2} \Biggr\} \cdot \T{n}{1} \, \diffo{\Gamma}.
\end{align}
\end{subequations}
}
\qed
\end{rmk}

The first term $ {\propto \gradx{\phi} \cdot \gradx{\phi}} $ of $ {Q_{\phi}} $ in Equation (\ref{eq:bulk.dissipation.alternative}) is always negative due to its negative sign, i.e., this term is a sink of energy. Although $ {\T{H}{2} : \T{H}{2}} $ is always positive, the function that multiplies it is indefinite if $ {\vartheta > 1} $. Otherwise, if $ {\vartheta < 1} $ this term is a sink of energy. The last term $ {\propto f(\phi) \, \T{H}{2} : \gradx{\phi} \otimes \gradx{\phi}} $ is indefinite since both $ {f(\phi)} $ and the Hessian of $ {\phi} $ are indefinite. Normalizing $ {\gradx{\phi}} $, this quadratic form $ {\T{H}{2} : \gradx{\phi} \otimes \gradx{\phi}} $ geometrically describes how the curvature in the $ {\gradx{\phi}} $ direction changes as we move along this direction, i.e., how the curvature changes across the interface.

\subsection*{\centering Interfacial Free Energy Exchange}

The exchanges (\ref{eq:general.budget}) in interfacial free energy can be expressed as
\begin{subequations}\label{eq:interfacial.dissipation}
\begin{align}
Q_s = & \int _{\Omega} \dfrac{Cn}{Lm} \left[ \T{H}{2} : \gradx{\T{j}{1}} - \T{D}{2} : \gradx{\phi} \otimes \gradx{\phi} \right] \, \diffo{\Omega}, \nonumber \\
= & - \int _{\Omega} \dfrac{Cn}{Lm} \left[ \dfrac{1}{PeLn}\T{H}{2} : \gradx{( M(\phi) \, \gradx{(\eta_{\phi}+\eta_s)} + \gradx{\eta_p} )} + \T{D}{2} : \gradx{\phi} \otimes \gradx{\phi} \right] \, \diffo{\Omega}, \\
S_s = & - \int _{\Gamma} \dfrac{Cn}{Lm} \left( \gradx{\phi} \cdot \gradx{\T{j}{1}} \right) \cdot \T{n}{1} \, \diffo{\Gamma}, \nonumber \\
= & \int _{\Gamma} \dfrac{1}{PeLn} \dfrac{Cn}{Lm} \left[ \gradx{\phi} \cdot \gradx{( M(\phi) \, \gradx{(\eta_{\phi}+\eta_s)} + \gradx{\eta_p} )} \right] \cdot \T{n}{1} \, \diffo{\Gamma}.
\end{align}
\end{subequations}

\emph{Physical interpretation}: The first term of the volumetric term $ {Q_s} $ has a similar meaning to that given for the kinetic and bulk free energy. The power expenditure carried out in an infinitesimal increment in the curvature of the phase field, $ {\diffo{\T{H}{2}}} $, done by the mass flux gradient is given by $ {\gradx{\T{j}{1}} : \diffo{\T{H}{2}}} $. Given an infinitesimal increment $ {\diffo{\T{H}{2}}} $, the local exchange of interfacial free energy is proportional to the mass flux gradient. That is, $ {Q_s} $ is the rate of work done by the mass flux gradient to achieve an increment in the curvature of the phase field. This interfacial free energy exchange acts as a source or a sink of energy.

We can interpret the first term in $ {Q_s} $ in a simple configuration. Suppose that the isosurface of $ {\phi} $ is locally flat, and that the mass flux varies only along the normal direction to the surface. Thus, the mass flux gradient leads to different number of particles diffusing at different points along the normal, therefore displacing the isosurfaces of $ {\phi} $. This in turn leads to a change in the magnitude of the gradient of $ {\phi} $, represented by the corresponding components of $ {\T{H}{2}} $, thus changing the interfacial free energy. Now suppose that the mass flux varies along the surface of constant $ {\phi} $. Thus, the mass flux gradient leads to a different number of diffusing particles along the tangential direction of the isosurface. This in turn induces a change in the curvature of the isosurface, thus changing the magnitude of the gradient of $ {\phi} $, which leads to a change in interfacial free energy. This is represented by the components of the Hessian of $ {\phi} $ describing the curvature of phase isosurfaces.

The second term in $ {Q_s} $ appears in the kinetic energy exchange. Even though it is discussed there, we interpret it here in the context of the interfacial energy. Looking at a small region on a phase field surface, if the normal velocity increases (decreases) in the direction of the isosurface normal (which is parallel to $ {\gradx{\phi}} $), then the interface widens (narrows), thereby decreasing (increasing) $ {\vert \gradx{\phi} \vert} $ thus becoming a sink (source) of interfacial energy. This term differs from that found in the kinetic energy exchange in scale and sign. As a consequence of the change of sign, if this term is a source (sink) of interfacial free energy, it will be a sink (source) of kinetic energy. The change in scaling expresses the relative importance of this term for different physical aspects of the energy budget of the Navier--Stokes--Cahn--Hilliard system.

The boundary term $ {S_s} $ represents the interfacial free energy diffusion on the boundaries by the mass flux.

\begin{rmk}\label{rmk:interfacial.dissipation}
Replacing the mass flux from Equation (\ref{eq:mass.flux}) into Equation (\ref{eq:interfacial.dissipation}) and using the Gauss divergence theorem we find where the source and sink of interfacial free energy come from. The volumetric $ {Q_s} $ and boundary $ {S_s} $ terms take the following forms in terms of $ {\phi} $:
\emph{
\begin{subequations}\label{eq:interfacial.dissipation.alternative}
\begin{align}
Q_s = & - \int _{\Omega} \Biggl\{ \dfrac{1}{PeLn} \dfrac{Cn}{Lm} \biggl[ - 2 (1 - 2\phi) \T{H}{2} : \gradx{\phi} \otimes \gradx{\phi} + \left( - 2 \phi (1 - \phi) + \dfrac{1}{2 \vartheta} \right) \T{H}{2} : \T{H}{2} \nonumber \\
+ & \dfrac{Cn}{Lm} \phi (1 - \phi) (\divx{\T{H}{2}}) \cdot (\divx{\T{H}{2}}) \biggr] + \dfrac{Cn}{Lm} \T{D}{2} : \gradx{\phi} \otimes \gradx{\phi} \Biggr\} \, \diffo{\Omega}, \\
S_s = & \int _{\Gamma} \dfrac{1}{PeLn} \dfrac{Cn}{Lm} \Biggl\{ \gradx{\phi} \cdot \biggl[ \gradx{\phi} \otimes \left( -2 (1 - 2\phi) \gradx{\phi} - \dfrac{Cn}{Lm} (1 - 2\phi) \divx{\T{H}{2}} \right) \nonumber \\
+ & \left( -2 \phi (1-\phi) + \dfrac{1}{2 \vartheta} \right) \T{H}{2} - \dfrac{Cn}{Lm} \phi (1 - \phi) \gradx{\divx{\T{H}{2}}} \biggr] + \dfrac{Cn}{Lm} \phi (1 - \phi) (\divx{\T{H}{2}}) \cdot \T{H}{2} \Biggr\} \cdot \T{n}{1} \, \diffo{\Gamma}.
% - \dfrac{Cn}{Lm} (1 - 2\phi) \T{H}{2} : \gradx{\phi} \otimes \divx{\T{H}{2}} + \dfrac{Cn}{Lm} (1 - 2\phi) \T{H}{2} : \gradx{\phi} \otimes \divx{\T{H}{2}}
\end{align}
\end{subequations}
}
\qed
\end{rmk}

There is only one term that does not change its sign, that is, the contraction on itself of $ {\divx{\T{H}{2}}} $. This term always acts as a sink. All the remaining terms that appear in $ {Q_s} $ in Equation (\ref{eq:interfacial.dissipation.alternative}) are indefinite, i.e., they may be either a source or a sink of interfacial free energy. The term $ {\propto f_1(\phi) \T{H}{2} : \T{H}{2} + f_2(\phi) \T{H}{2} : \gradx{\phi} \otimes \gradx{\phi}} $ also appears in the bulk free energy exchange $ {Q_{\phi}} $, with the same sign and scales. Thus, its physical meaning is exactly the same.

\subsection*{ \centering Potential Energy Exchange}

Finally, the forms of the terms in (\ref{eq:general.budget}) for the potential energy are
\begin{subequations}\label{eq:potential.dissipation}
\begin{align}
Q_p = & \int _{\Omega} \dfrac{Bo}{We} \, \T{e}{1}_2 \cdot \T{j}{1} \, \diffo{\Omega} + \int _{\Omega} \dfrac{Bo}{We} \phi \, v_2 \, \diffo{\Omega}, \nonumber \\
= & - \int _{\Omega} \dfrac{1}{PeLn} \dfrac{Bo}{We} \T{e}{1}_2 \cdot \left[ M(\phi) \, \gradx{(\eta_{\phi}+\eta_s)} + \gradx{\eta_p} \right] \diffo{\Omega} + \int _{\Omega} \dfrac{Bo}{We} \phi \, v_2 \, \diffo{\Omega}, \\
S_p = & - \int _{\Gamma}\dfrac{Bo}{We} x_2 \, \T{j}{1} \cdot \T{n}{1} \, \diffo{\Gamma}, \nonumber \\
= & \int _{\Gamma} \dfrac{1}{PeLn} \dfrac{Bo}{We} x_2 \left[ M(\phi) \, \gradx{(\eta_{\phi}+\eta_s)} + \gradx{\eta_p} \right] \cdot \T{n}{1} \, \diffo{\Gamma}.
\end{align}
\end{subequations}

\emph{Physical interpretation}: The sources and sinks of potential energy described by $ {Q_p} $ are related to relative motion in the $ {\T{e}{1}_2} $ direction. The first term describes the diffusive flux in this direction and therefore can be a source or sink of potential energy. This contribution acts as a source -- if the flux points in the same direction as $ {\T{e}{1}_2} $ -- since it leads to fluid particles diffusing against the pull of gravity (pointing towards -$ {\T{e}{1}_2} $) thereby gaining potential energy. This contribution acts as a sink of energy if the direction of the mass flux is parallel to gravity, thus accepting the pull and losing potential energy. The second term is related to the non-diffusive motion of the fluid particles. If the velocity points towards $ {\T{e}{1}_2} $ then the particle moves against gravity, gaining potential energy, and vice versa. This term cancels out the third term appearing in the kinetic energy exchanges and therefore it does not play any role in the energy budget equation.

\begin{rmk}\label{rmk:potential.dissipation}
As we did in remarks \ref{rmk:bulk.dissipation} and \ref{rmk:interfacial.dissipation}, we obtain explicit expressions in term of $ {\phi} $ for the volumetric $ {Q_p} $ and boundary $ {S_p} $ terms:
\emph{
\begin{subequations}\label{eq:potential.dissipation.alternative}
\begin{align}
Q_p = & - \int _{\Omega} \dfrac{1}{PeLn} \dfrac{Bo}{We} \biggl[ \left( -2\phi (1-\phi) + \dfrac{1}{2\vartheta} \right) \gradx{\phi} \cdot \T{e}{1}_2 \nonumber \\
+ & \dfrac{Cn}{Lm} (1 - 2\phi) \T{H}{2} : \gradx{\phi} \otimes \T{e}{1}_2 + \dfrac{Bo}{We} \biggr] \diffo{\Omega} + \int _{\Omega} \dfrac{Bo}{We} \phi \, v_2 \, \diffo{\Omega}, \\
S_p = & \int _{\Gamma} \dfrac{1}{PeLn} \dfrac{Bo}{We} \Biggl\{ \dfrac{Cn}{Lm} \phi (1-\phi) \T{e}{1}_2 \cdot \T{H}{2} \nonumber \\
+ & x_2 \left[ \left( -2 \phi (1-\phi) + \dfrac{1}{2 \vartheta} \right) \gradx{\phi} - \dfrac{Cn}{Lm} \phi (1-\phi) \divx{\T{H}{2}} + \dfrac{Bo}{We} \T{e}{1}_2 \right] \Biggr\} \cdot \T{n}{1} \, \diffo{\Gamma}.
\end{align}
\end{subequations}
}
\qed
\end{rmk}

Despite identifying at least one negative definite term (sink of energy) in the energy exchanges discussed above, the definiteness of the potential energy exchanges cannot be determined beforehand. Finally, normalizing $ {\gradx{\phi}} $ the term $ {\T{H}{2} : \gradx{\phi} \otimes \T{e}{1}_2} $ geometrically describes how the curvature in the $ {\gradx{\phi}} $ direction changes as we move along the $ {\T{e}{1}_2} $ direction.

\subsection*{\centering Total Energy Budget}

After detailing the different energy exchange terms in the system, we address the total energy exchanges in the system. Integrating the source and sink terms along with the fluxes over time leads to the total energy exchange terms, where the volumetric terms are defined as
\begin{equation}
\mathcal{E}_{\jmath} = \int_{t} Q_{\jmath} \, \diffo{t} = \int_{t} \int_{\Omega} \epsilon_{\jmath} \, \diffo{\Omega} \diffo{t}, \qquad \jmath = \{k,\phi,s,p\},
\end{equation}
and the input/output of energy diffused across the boundaries is
\begin{equation}
\mathcal{E}_{\jmath}^{io}\Big|_{\textit{diff}} = \int_{t} S_{\jmath} \, \diffo{t} = \int_{t} \int_{\Gamma} \epsilon_{\jmath}^{io}\Big|_{\textit{diff}} \, \diffo{\Gamma} \diffo{t}, \qquad \jmath = \{k,\phi,s,p\}.
\end{equation}
If there exists a momentum flux across the boundaries, the energy must be advected according to
\begin{equation}
\mathcal{E}_{\jmath}^{io}\Big|_{\textit{adv}} = - \int_{t} \int_{\Gamma} e_{\jmath} \, \T{v}{1} \cdot \T{n}{1} \, \diffo{\Gamma} \diffo{t}, \qquad \jmath = \{k,\phi,s,p\},
\end{equation}
and the sum of $ \mathcal{E}_{\jmath}^{\textit{io}}\Big|_{\textit{diff}} + \mathcal{E}_{\jmath}^{\textit{io}}\Big|_{\textit{adv}} $ yields the total input/output of energy across the boundaries.

Finally, the energy budget is defined as 
\begin{equation}
E_k + E_{\phi} + E_s + E_p - (\mathcal{E}_k + \mathcal{E}_{\phi} + \mathcal{E}_s + \mathcal{E}_p + \mathcal{E}_k^{\textit{io}} + \mathcal{E}_{\phi}^{\textit{io}} + \mathcal{E}_s^{\textit{io}} + \mathcal{E}_p^{\textit{io}}) = C,
\end{equation}
for an arbitrary constant $ {C} $. We analyze the energy exchange over the whole domain and use no-flux boundary conditions for all fields in the numerical examples discussed in the following. Under such setup, the energy budget equation reads
\begin{equation}\label{eq:energy.budget}
\sum _{\jmath = \{k,\phi,s,p\}} (E_{\jmath} - \mathcal{E}_{\jmath})(t) = E_k + E_{\phi} + E_s + E_p - (\mathcal{E}_k + \mathcal{E}_{\phi} + \mathcal{E}_s + \mathcal{E}_p) = C.
\end{equation}
As mentioned before, we relate $ {E_{\jmath}} $ and $ {\mathcal{E}_{\jmath}} $, $ {\jmath = \{k,\phi,s,p\}} $, to extensive energies and their energy exchanges, whereas $ {e_{\jmath}} $ and $ {\epsilon_{\jmath}} $ are associated to intensive energies and their energy exchanges, respectively.

\textbf{Remarks \ref{rmk:bulk.dissipation} -- \ref{rmk:potential.dissipation}} show which terms work as sinks or as both sources/sinks of energy, as functions of the phase field. Concluding, the sum of energy exchanges in terms of the phase field reads
\begin{equation}\label{eq:dissipations.volume}
\begin{split}
\vphantom{\sum _0^0} \sum _{\jmath = \{k,\phi,s,p\}} \!\!\! \epsilon_{\jmath} = & \dfrac{Cn(Lm-We)}{WeLm} \, \T{D}{2} : \gradx{\phi} \otimes \gradx{\phi} - \dfrac{2c}{Re} \T{D}{2} : \T{D}{2} \\
\vphantom{\sum _0^0} - & \dfrac{\phi (1-\phi)}{PeLn} \left( -2 + \dfrac{1}{2 \vartheta \phi (1-\phi)} \right)^2 \gradx{\phi} \cdot \gradx{\phi} \\
\vphantom{\sum _0^0} - & \dfrac{\phi (1 - \phi)}{PeLn} \left( \dfrac{Cn}{Lm} \right)^2 (\divx{\T{H}{2}}) \cdot (\divx{\T{H}{2}}) \\
\vphantom{\sum _0^0} - & \dfrac{2}{PeLn} \dfrac{Cn}{Lm} \left( -2 \phi (1-\phi) + \dfrac{1}{2 \vartheta} \right) \T{H}{2} : \T{H}{2} \\
\vphantom{\sum _0^0} + & \dfrac{4}{PeLn} \dfrac{Cn}{Lm} (1 - 2 \phi) \, \T{H}{2} : \gradx{\phi} \otimes \gradx{\phi} \\
\vphantom{\sum _0^0} - & \dfrac{1+\phi (1-\phi)}{PeLn} \dfrac{Bo}{We} \left( -2 + \dfrac{1}{2 \vartheta \phi (1-\phi)} \right) \gradx{\phi} \cdot \T{e}{1}_2 \\
\vphantom{\sum _0^0} - & \dfrac{1}{PeLn} \dfrac{Bo}{We} \left[ (1 - 2\phi) \dfrac{Cn}{Lm} \T{H}{2} : \gradx{\phi} \otimes \T{e}{1}_2 + \dfrac{Bo}{We} \right].
\end{split}
\end{equation}

In the system governed by Navier--Stokes--Cahn--Hilliard equations, there are three volumetric terms that always work as energy sinks. They are: $ {\propto f_1(\phi) \T{D}{2} : \T{D}{2}} $, $ {\propto f_2(\phi) \gradx{\phi} \cdot \gradx{\phi}} $ and $ {\propto f_3(\phi) (\divx{\T{H}{2}}) \cdot (\divx{\T{H}{2}}}) $. If the critical temperature ratio $ {\vartheta} $ is less than one, the term $ {\propto f_4(\phi) \, \T{H}{2} : \T{H}{2}} $ is also a sink. The functions $ {f_1} $, $ {f_2} $, $ {f_3} $ and $ {f_4} $ are stated as
\begin{subequations}
\begin{align}
f_1(\phi) = & - \dfrac{2c}{Re} \\
f_2(\phi) = & - \dfrac{\phi (1-\phi)}{PeLn} \left( -2 + \dfrac{1}{2 \vartheta \phi (1-\phi)} \right)^2 \\
f_3(\phi) = & - \dfrac{1}{PeLn} \left( \dfrac{Cn}{Lm} \right)^2 \phi (1-\phi) \\
f_4(\phi) = & - 2 \dfrac{1}{PeLn} \dfrac{Cn}{Lm} \left( -2 \phi (1-\phi) + \dfrac{1}{2 \vartheta} \right).
\end{align}
\end{subequations}
Unfortunately, we cannot establish beforehand if the remaining terms work as sources or sinks of energy, as these terms are indefinite.

\begin{figure}[!t]
\centering
  \includegraphics[width=0.3\textwidth]{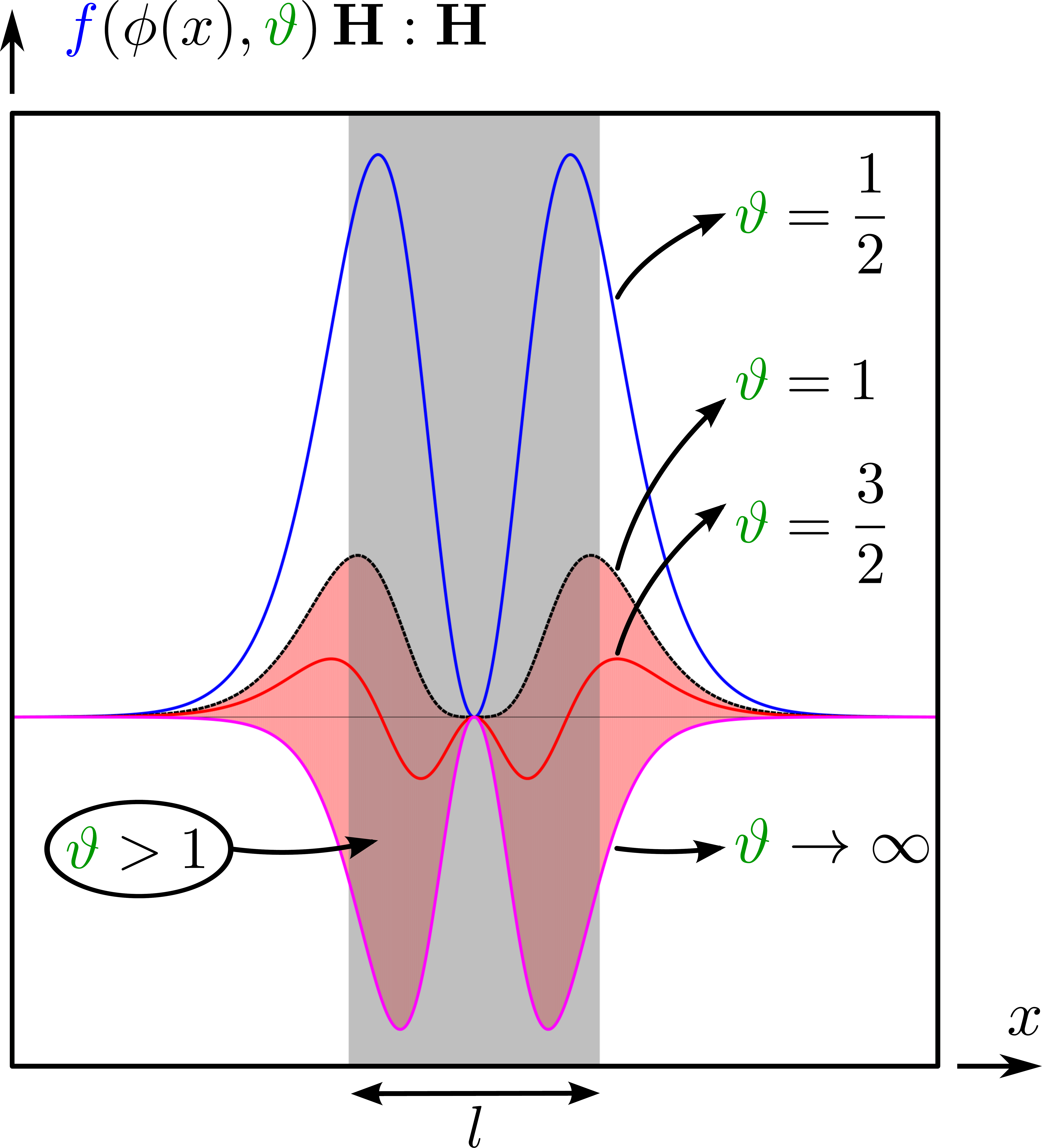}
  \caption{(color online) The plot depicts the function that appears in both the bulk and the interfacial free energies, $ {f(\phi,\vartheta) \, \mathbf{H} : \mathbf{H} = [-2 \phi (1-\phi) + \frac{1}{2 \vartheta}] \, \mathbf{H} : \mathbf{H}} $ along the $ {\phi} $ axis for $ {\vartheta} $ from $ {0.5} $ to $ {\infty} $. The phase field is defined by a hyperbolic tangent, $ {\phi(x)=0.95[0.5 \, \text{tanh} (10x)+0.5]+0.025} $. The grey region represents the interface length $ {l} $, whereas the red region depicts the region where this function lives when the phase segregation is assumed, i.e., $ {\vartheta > 1} $.}
\label{fg:function.H:H}
\end{figure}

Figure \ref{fg:function.H:H} depicts the behavior of the term $ {f(\phi(x),\vartheta) \, \T{H}{2} : \T{H}{2} = [-2 \phi (1-\phi) + \frac{1}{2 \vartheta}] \, \mathbf{H} : \mathbf{H}} $ that appears in both the bulk and interfacial free energies, along the $ {\phi} $ axis for $ {\vartheta} $ from $ {0.5} $ to $ {\infty} $. The phase field is defined by a hyperbolic tangent, $ {\phi(x)=0.95[0.5 \tanh (10x)+0.5]} $ $ {+0.025} $. The grey region represents the interface length, whereas the red region depicts the region where this function lives when phase segregation takes place, i.e., $ {\vartheta > 1} $. The term $ {f(\phi(x),\vartheta) \, \mathbf{H} : \mathbf{H}} $ is always positive if $ {\vartheta < 1} $. After scaling by $ {- 2 \frac{1}{PeLn} \frac{Cn}{Lm}} $, the energy exchange term becomes negative definite, thus it is a sink of energy. However, if $ {\vartheta > 1} $ (red region depicted in Figure \ref{fg:function.H:H}), the energy exchange term is indefinite, being a source of energy in the middle region of the interface and becoming a sink of energy as we move away from the interface. When the phase is segregated, far away from the interfaces, this term does not play any role. In the extreme case, $ {\vartheta \rightarrow \infty} $, this term acts as a source of energy.

Finally, the sum of boundary sources and sinks of energies, as functions of the phase field, reads
\begin{equation}\label{eq:dissipations.boundary}
\begin{split}
\vphantom{\sum _0^0} \sum _{\jmath = \{k,\phi,s,p\}} \!\!\! \epsilon^{io}_{\jmath}\Big|_{\textit{diff}} = & \Biggl\{ \T{v}{1} \cdot \left( \dfrac{2c}{Re} \T{D}{2} - \dfrac{p}{Re} \T{1}{2} - \dfrac{Cn}{We} \, \gradx{\phi} \otimes \gradx{\phi} \right) \\
\vphantom{\sum _0^0} + & \dfrac{1}{PeLn} \left( \dfrac{1}{2\vartheta} \ln \dfrac{\phi}{1-\phi} + 1 - 2\phi \right) \Biggl[ \left( -2 \phi (1-\phi) + \dfrac{1}{2\vartheta} \right) \gradx{\phi} \\
\vphantom{\sum _0^0} - & \dfrac{Cn}{Lm} \phi (1-\phi) \, \divx{\T{H}{2}} + \dfrac{Bo}{We} \T{e}{1}_2 \Biggr] + \dfrac{1}{PeLn} \dfrac{Cn}{Lm} \left( -2 \phi (1-\phi) + \dfrac{1}{2 \vartheta} \right) \gradx{\phi} \cdot \T{H}{2} \\
\vphantom{\sum _0^0} - & \dfrac{1}{PeLn} \dfrac{Cn}{Lm} \, \gradx{\phi} \cdot \biggl[ \gradx{\phi} \otimes \left( 2 (1 - 2\phi) \gradx{\phi} + \dfrac{Cn}{Lm} (1 - 2\phi) \divx{\T{H}{2}} \right) \\
\vphantom{\sum _0^0} - & \left( -2 \phi (1-\phi) + \dfrac{1}{2 \vartheta} \right) \T{H}{2} + \dfrac{Cn}{Lm} \phi (1 - \phi) \gradx{\divx{\T{H}{2}}} \biggr] \\
\vphantom{\sum _0^0} + & \dfrac{\phi (1 - \phi)}{PeLn} \left( \dfrac{Cn}{Lm} \right)^2 (\divx{\T{H}{2}}) \cdot \T{H}{2} + \dfrac{\phi (1-\phi)}{PeLn} \dfrac{Bo}{We} \dfrac{Cn}{Lm} \T{e}{1}_2 \cdot \T{H}{2} \\
\vphantom{\sum _0^0} + & \dfrac{x_2}{PeLn} \dfrac{Bo}{We} \left[ \left( -2 \phi (1-\phi) + \dfrac{1}{2 \vartheta} \right) \gradx{\phi} - \phi (1-\phi) \dfrac{Cn}{Lm} \divx{\T{H}{2}} + \dfrac{Bo}{We} \T{e}{1}_2 \right] \Biggr\} \cdot \T{n}{1}.
\end{split}
\end{equation}

In our numerical experiments, we present the energy terms using Equations (\ref{eq:energy.k})-(\ref{eq:energy.p}) and their related energy exchange terms using Equations (\ref{eq:kinetic.dissipation}), (\ref{eq:bulk.dissipation}), (\ref{eq:interfacial.dissipation}) and (\ref{eq:potential.dissipation}). We have opted to compute energy exchanges using the expressions containing the mass flux as a variable, in those terms related to bulk, interfacial and potential energies, i.e., Equations (\ref{eq:bulk.dissipation}), (\ref{eq:interfacial.dissipation}) and (\ref{eq:potential.dissipation}) instead of Equations (\ref{eq:bulk.dissipation.alternative}), (\ref{eq:interfacial.dissipation.alternative}) and (\ref{eq:potential.dissipation.alternative}) presented in \textbf{Remarks \ref{rmk:bulk.dissipation} -- \ref{rmk:potential.dissipation}}. The reasons to opt for this approach are twofold. First, we take advantage of the mixed formulation of Cahn--Hilliard Equation, since the auxiliary variable is the chemical potential. Second, at most second-order derivatives are required for the phase field, whereas Equations (\ref{eq:bulk.dissipation.alternative}), (\ref{eq:interfacial.dissipation.alternative}) and (\ref{eq:potential.dissipation.alternative}) require fourth-order derivatives of the phase field, which leads to lower order approximations of the fields of interest.

To simplify the exposition of results, we split the volumetric energy exchange terms. Three terms appear in the kinetic energy exchanges: the power expenditure done by the viscous stress $ {\epsilon_k^{visc} = - \T{D}{2} : \T{T}{2}^{visc}} $, the power expenditure done by the capillary stress $ {\epsilon_k^{s} = - \T{D}{2} : \T{T}{2}^{s}} $, and a term related to the Boussinesq approximation $ {\epsilon_k^{buoy} = - \frac{Bo}{We} \phi \, v_2} $. The bulk free energy exchange is described by only one term, that is, the power expenditure done by the mass flux on the bulk free energy gradient, $ {\epsilon_{\phi} = \gradx{\eta_{\phi}} \cdot \T{j}{1}} $. The interfacial free energy exchange is split in a term related to changes in the curvature done by the mass flux gradient $ {\epsilon_{s}^{curv} = \frac{Cn}{Lm} \T{H}{2} : \gradx{\T{j}{1}}} $ and a term that relates the strain rate and capillary effects, $ {\epsilon_{s}^{s} = - \frac{Cn}{Lm} \T{D}{2} : \gradx{\phi} \otimes \gradx{\phi}} $, similarly to $ {\epsilon_k^{s}} $. Finally, in the potential energy exchange, the first term describes the mass flux in the vertical direction $ {\epsilon_p^{mass} = \frac{Bo}{We} \, \T{e}{1}_2 \cdot \T{j}{1}} $, while the second term $ {\epsilon_p^{buoy} = - \epsilon_k^{buoy} = \frac{Bo}{We} \phi \, v_2} $, which also appears in the kinetic energy exchange, is related to the Boussinesq approximation. The same notation is employed for extensive quantities. Finally, we employ the trapezoidal rule to integrate over time the energy exchanges.

%===============================================================================================================%

\section{Numerical Scheme: Divergence Conforming B-spline Spaces}\label{sec:num.scheme}

Isogemeotric analysis has been used successfully to solve high-order phase-field models, including the Cahn--Hilliard equation \citep{GOM08,VIG13}, the Navier--Stokes--Korteweg equations \citep{GOM10}, the Swift-Hohenberg equation \citep{GOM12a} and the phase-field crystal equation \citep{GOM12b,VIG13,VIG15b}. Stability conditions in models for electromagnetism such as the Maxwell equations \citep{BUF10a} and flow models including Stokes \citep{BUF10b,EVA13a}, and Navier--Stokes equations \citep{EVA13b}, have been solved exactly using curl- and divergence-conforming spaces, respectively. Solving incompressible flow models using divergence-conforming spaces produces discrete pointwise divergence-free velocity fields. The advantages of such fields in the conservation of kinetic energy, vorticity, enstrophy, and helicity are discussed by \citep{EVA13c} for the Navier--Stokes equations and are extensibly exploited herein.

The discrete model is solved using the PetIGA-MF \citep{SAR15,VIG15a}, a high-performance framework built on top of PetIGA \citep{COR14,DAL15}, which uses structure-preserving B-spline basis functions. This framework simplifies the solution of systems of high-order partial differential equations, where multifield strategies can provide high order of approximation and smoothness in the basis functions, as well as structure-preserving discretizations that allow exact satisfaction of discrete stability conditions \citep{BUF11,COR15}.

The idea of structure-preserving spaces is based on the satisfaction of the exact sequence given by the discrete version the de Rham diagram \citep{BUF11}. Here, two spaces are said to be conforming to an operator if they satisfy a step in the sequence that corresponds to that particular operator. PetIGA-MF admits multifield discretizations and provides gradient-, curl-, divergence- and integral-conforming discrete spaces, allowing the user to discretize a specific problem in a stable manner.

We use a mixed formulation of the Cahn--Hilliard equation, taking the chemical potential $ {\eta} $ as an auxiliary variable. We do this to reduce the computational cost of using high order and high continuity basis functions \citet{COL12a,COL12b,COL14}, and to avoid complications with the imposition of nonlinear boundary conditions that arise from the discretization of the mass flux in the primal form. Thus, we have four variables, the velocity, pressure, phase, and chemical potential fields. We discretize the velocity and pressure variables using a divergence- and integral-conforming conjugated pair of spaces that satisfy the inf-sup stability condition exactly and render a pointwise divergence-free discrete velocity field. The phase field and chemical potential variables are discretized using $ {H^1} $ spaces. No penetration and free-slip boundary conditions are imposed on the velocity since we are not seeking to study boundary layer effects in this work. In addition, homogeneous Neumann boundary conditions are applied to all the other variables.

We employ the Generalized-$ {\alpha} $ time integrator presented by \citet{CHU93,JAN00,ESP15b} to advance all fields in time. In all examples presented here, we employ a spectral radius, $ {\rho_\infty} $, equal to $ {0.9} $. This means that we add a small amount of numerical dissipation in the system to avoid numerical instabilities. We also use time-step adaptivity based on keeping the local truncation error of the method under a prescribed tolerance.

%===============================================================================================================%

\section{Problem setup}\label{sec:problem}

We perform four numerical experiments to investigate the energy exchanges in the Navier--Stokes--Cahn--Hilliard system. In addition, we analyze the main features of this kind of flow as well as the time evolution of the half distance between the meniscus, i.e., the bridge length as defined in \citet{EGG99}. The first two-dimensional simulation (case $ {\#1} $) is performed without buoyancy effects, whereas the second and third two-dimensional simulations (cases $ {\#2} $ and $ {\#3} $) take into account buoyancy effects to analyze the merging of droplets when they are rising. The fourth simulation, which is three-dimensional (case $ {\#4} $), includes buoyancy effect and deals with a rising droplet that coalesces with a film of fluid.

Table \ref{tb:physical.parameters} lists (in the first column, case $ {\#1} $) the physical parameters employed in the simulation without buoyancy and (in the second and third columns, cases $ {\#2} $ and $ {\#3} $, respectively) the physical parameters employed in each simulation that takes into account buoyancy. Cases $ {\#2} $ and $ {\#3} $ differ only in the dimensionless $ {Cn/We} $ ratio. We employ $ {(Cn/We)^{-1} = 10^2} $ and $ {10^3} $ for those simulations, respectively. The fourth column of Table \ref{tb:physical.parameters} lists the parameters used in case $ {\#4} $.

\begin{figure}[!t]
\centering
  \includegraphics[width=0.55\textwidth]{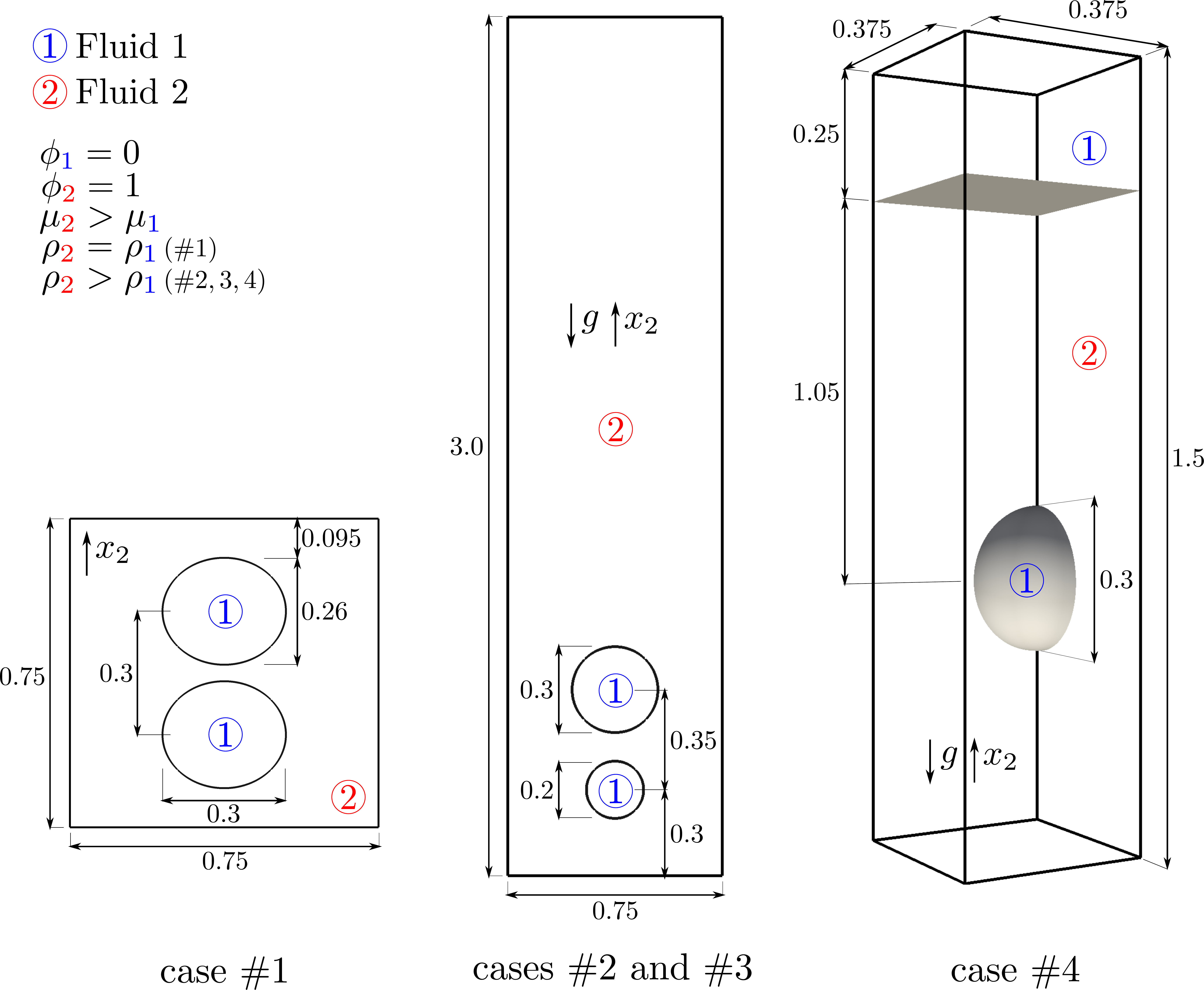}
\caption{Initial condition in the 2D domain (on the left) without buoyancy effects, case $ {\#1} $; in the 2D domain (in the middle) with buoyancy effects, cases $ {\#2} $ and $ {\#3} $; and in the 3D domain (on the right) with buoyancy effects, case $ {\#4} $.}
\label{fg:domain}
\end{figure}

The domains we consider are shown in Figure \ref{fg:domain}, where the vertical axis is the $ {x_2} $ direction. On the left, Figure \ref{fg:domain} shows the two-dimensional domain and the initial location of the droplets employed in case $ {\#1} $. In the middle, Figure \ref{fg:domain} depicts the domain and the initial location of the droplets employed in cases $ {\#2} $ and $ {\#3} $. On the right, Figure \ref{fg:domain} shows the domain with the initial configuration for the three-dimensional case $ {\#4} $. Due to symmetry, only a quarter of the domain is considered in case $ {\#4} $.

Table \ref{tb:mesh} lists (from the first to the third column) the number of mesh nodes used in the two-dimensional cases $ {\#1} $, $ {\#2} $, and $ {\#3} $, respectively. The last column lists the number of mesh nodes employed in the three-dimensional case $ {\#4} $. Table \ref{tb:discrete.spaces} shows the polynomial degree and continuity employed. All meshes are uniform.

In the first simulation, case $ {\#1} $, the droplets and the interstitial fluid are endowed with the same density, but the droplet is less viscous than the interstitial fluid. In the remaining cases, cases $ {\#2} $ to $ {\#4} $, droplets are lighter and less viscous (endowed with density, $ {\rho_1} $, and viscosity, $ {\mu_1} $) than the interstitial fluid (endowed with a density, $ {\rho_2} $, and viscosity, $ {\mu_2} $). To represent the phobic interections between the phases, $ {\vartheta = 3/2 > 1} $ is employed.

Finally, the initial condition in case $ {\#1} $ is given by
\begin{gather*}
h = 0.01; \quad b_1 = 1.0; \quad b_2 = 0.75 \\
a_1 = \tanh \left\{ \left[ \left( \dfrac{x_1-0.375}{b_1} \right)^2 + \left( \dfrac{x_2-0.375-0.15}{b_2} \right)^2 - 0.15^2 \right] \dfrac{h^{-1}}{0.3 + h} \right\} 0.499-0.5; \\
a_2 = \tanh \left\{ \left[ \left( \dfrac{x_1-0.375}{b_1} \right)^2 + \left( \dfrac{x_2-0.375+0.15}{b_2} \right)^2 - 0.15^2 \right] \dfrac{h^{-1}}{0.3 + h} \right\} 0.499-0.5; \\
\phi (x_1,x_2,t=0) = a_1(x_1,x_2) + a_2(x_1,x_2) + 1,
\end{gather*}
while in cases $ {\#2} $ and $ {\#3} $ they are
\begin{gather*}
h = 0.01; \\
a_1 = \tanh \left\{ \left[ (x_1-0.375)^2 + (x_2-0.65)^2 - 0.15^2 \right] \dfrac{h^{-1}}{0.3 + h} \right\} 0.499-0.5; \\
a_2 = \tanh \left\{ \left[ (x_1-0.375)^2 + (x_2-0.30)^2 - 0.10^2 \right] \dfrac{h^{-1}}{0.2 + h} \right\} 0.499-0.5; \\
\phi (x_1,x_2,t=0) = a_1(x_1,x_2) + a_2(x_1,x_2) + 1.
\end{gather*}
In case $ {\#4} $, the initial condition is defined as
\begin{gather*}
h = 0.02; \\
a_1 = \tanh \left\{ \left[ (x_1-0.375)^2 + (x_2-0.5)^2 + (x_3-0.375)^2 - 0.15^2 \right] \dfrac{h^{-1}}{0.3 + h} \right\} 0.499-0.5; \\
a_2 = - \tanh \left( \dfrac{x_2-1.25}{h} \right) 0.499+0.5; \\
\phi (x_1,x_2,x_3,t=0) = a_1(x_1,x_2,x_3) + a_2(x_2) + 1.
\end{gather*}

\begin{table*}[!t]
  \caption{Physical Parameters}
  \label{tb:physical.parameters}
  \centering
  \footnotesize
  \begin{tabular}{c c c c c c}
  \toprule
  Number      &   & case $ {\#1} $, 2D       & case $ {\#2} $ 2D      & case $ {\#3} $, 2D         & case $ {\#4} $, 3D \\ %[0.5cm]
  \midrule
  $ {PeLn} $  & - & $ {10^3} $               & $ {10^3} $             & $ {10^3} $                 & $ {10^3} $ \\ %[0.5cm]
  $ {Re} $    & - & $ {10^3} $               & $ {10^3} $             & $ {10^3} $                 & $ {10^3} $ \\ %[0.5cm]
  $ {Bo/We} $ & - & $ {1} $                  & $ {1} $                & $ {1} $                    & $ {1} $ \\ %[0.5cm]
  $ {(Cn/Lm)^{-1}} $ & - & $ {2 \cdot 10^4} $       & $ {2 \cdot 10^4} $     & $ {2 \cdot 10^4} $         & $ {10^4} $ \\ %[0.5cm]
  $ {(Cn/We)^{-1}} $ & - & $ {10^3} $               & $ {\bt{10^2}} $        & $ {\bt{10^3}} $            & $ {10^3} $ \\ %[0.5cm]
  \midrule
  \multicolumn{6}{c}{$ {\mu_2/\mu_1 = 10} $; $ {\vartheta = 3/2} $ (all cases)} \\
  \bottomrule
  \end{tabular}
\end{table*}
\begin{table*}[!t]
  \caption{Meshes}
  \label{tb:mesh}
  \centering
  \footnotesize
  \begin{tabular}{c c c c c c}
  \toprule
                                         &   & case $ {\#1} $, 2D       & case $ {\#2} $, 2D       & case $ {\#3} $, 2D       & case $ {\#4} $, 3D \\ %[0.5cm]
  \midrule
  $ {(n_{x_{1}},n_{x_{2}},n_{x_{3}})} $  & - & $ {(256,256)} $          & $ {(256,1024)} $         & $ {(256,1024)} $         & $ {(96,384,96)} $ \\ %[0.5cm]
  \bottomrule
  \end{tabular}
\end{table*}
\begin{table*}[!t]
  \caption{Discrete Spaces}
  \label{tb:discrete.spaces}
  \centering
  \footnotesize
  \begin{tabular}{c c c c}
  \toprule
                  &   & $ {v_i} $                               & $ {p,\phi,\eta} $ \\ %[0.5cm]
  \midrule
  $ {(p,k)_{d}} $ & - & $ {(3,2)_{x_{i}},(2,1)_{x_{j \ne i}}} $ & $ {(2,1)} $       \\ %[0.5cm]
  \midrule
  \multicolumn{4}{c}{$ {p} $ - degree; $ {k} $ - continuity; $ {d} $ - direction} \\
  \bottomrule
  \end{tabular}
\end{table*}

%===============================================================================================================%

\section{Numerical Investigations}\label{sec:results}

In this section, we present the four numerical experiments described in the previous section. These use simple domains and focus on the physical aspects of the energy exchanges. The first simulation deals with the merging of droplets without gravitational effects in two dimensions. The second and third simulations deal with rising droplets that merge as they evolve. The fourth and final simulation is three-dimensional, and deals with a single droplet, rising to merge with a thin film of fluid. In the first simulation, the droplets are less viscous than the interstitial fluid and both fluids are endowed with the same density. In the remaining simulations, the droplets are lighter and less viscous than the interstitial fluid.

\subsection*{\centering Two-Dimensional Investigation: Case $ {\#1} $}

In the absence of gravity, the flow is driven by surface tension and liquid droplets in close proximity immersed in a liquid continuum merge into a single bigger droplet.

\subsubsection{General Flow Features in Droplet Dynamics: $ {\#1} $}

\begin{figure}[!t]
\centering
  \includegraphics[width=0.65\textwidth]{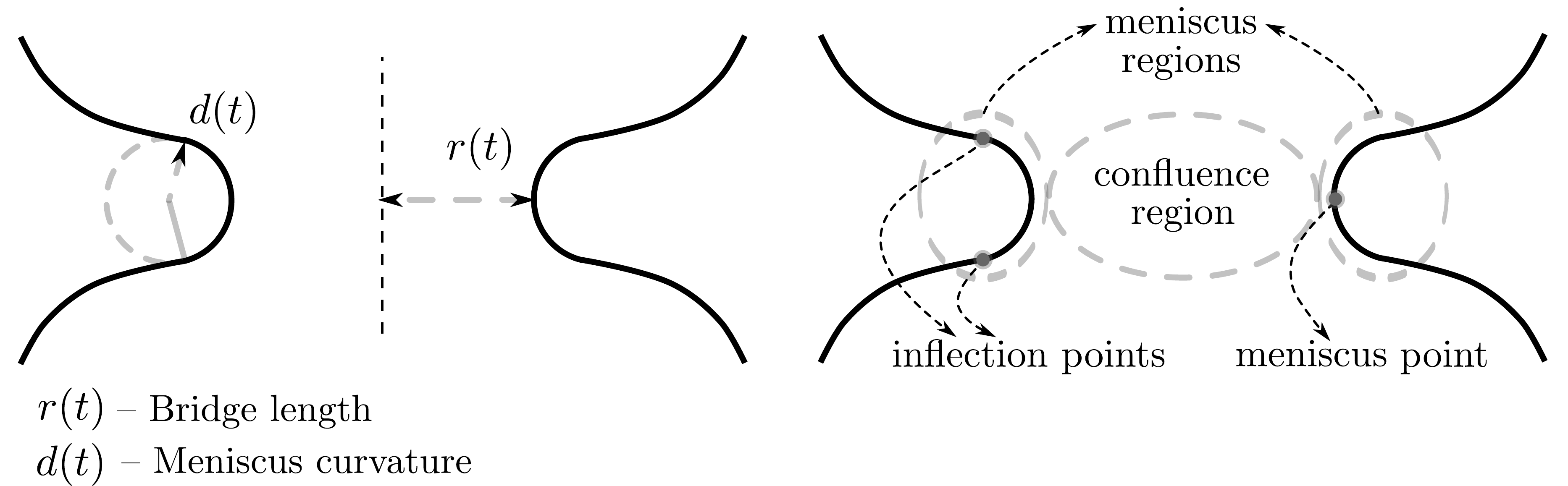}
  \caption{A schematic representation of the meniscus and confluence regions defined over the isocurve $ {\phi = 0.5} $ during the merging of droplets.}
\label{fg:meniscus.confluence.regions}
\end{figure}

Let us define two characteristic regions in the merging of droplets. The first one is the region where the interface of the droplets breaks down to combine them into a single entity. This region does not contain the interface, only the region between the broken interfaces. We call it the confluence region. The second one is the broken interface, which has the highest curvature. We call it the meniscus region. These regions are depicted on the right of Figure \ref{fg:meniscus.confluence.regions}. On the left of this Figure, we depict the Bridge Length and the Meniscus Curvature, following the usual definition by \citet{EGG99}.

\begin{figure}[!t]
\centering
  \includegraphics[width=0.7\textwidth]{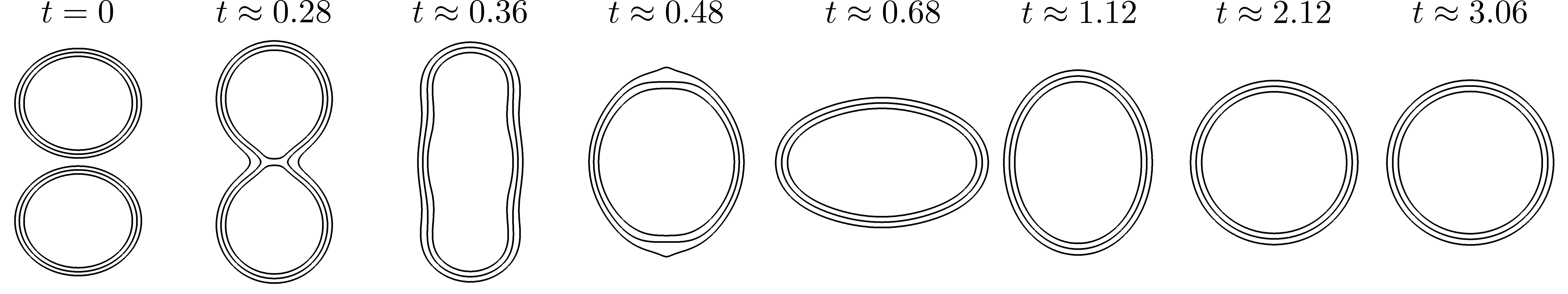}
  \caption{Isocurves for $ {\phi = 0.1} $, $ {0.5} $ and $ {0.9} $.}
\label{fg:merging.no.buoyancy}
\end{figure}

Initially, both droplets are elliptical. To achieve a state that requires a minimum energy, they merge into a bigger droplet, which minimizes the surface area. This results in a circular droplet. Figure \ref{fg:merging.no.buoyancy} shows isocurves of the phase-field evolution, being the initial state, two elliptical droplets and the final one, a steady bigger circular droplet.

\subsubsection{Meniscus and Bridge Evolution: $ {\#1} $}

According to the theory presented by \citet{EGG99}, at early stages of the merging process, the bridge length (half distance between the menisci) scales with $ {\sim t \ln t} $, i.e., $ {r(t) \propto t \ln t} $ and it holds up to $ {r(t) \lesssim 0.03 R} $, where $ {t} $ is the time counted from the initial contact between the droplets, and $ {R} $ being the droplet radius. In \citet{EGG99}, the authors used the Stokes system together with the sharp interface method to track the interface and model the viscous motion. If $ {r \gtrsim 0.03 R} $ the scaling law changes and the bridge length scales with $ {\sim \sqrt{t}} $, i.e., $ {r(t) \propto \sqrt{t}} $. Although the theoretical scaling law at early stages of the merging is $ {\sim t \ln t} $, a linear growth has been reported by \citet{THO05} and \citet{AAR05} in experimental analyses.

\begin{figure}[!t]
\centering
  \includegraphics[height=0.3\textheight]{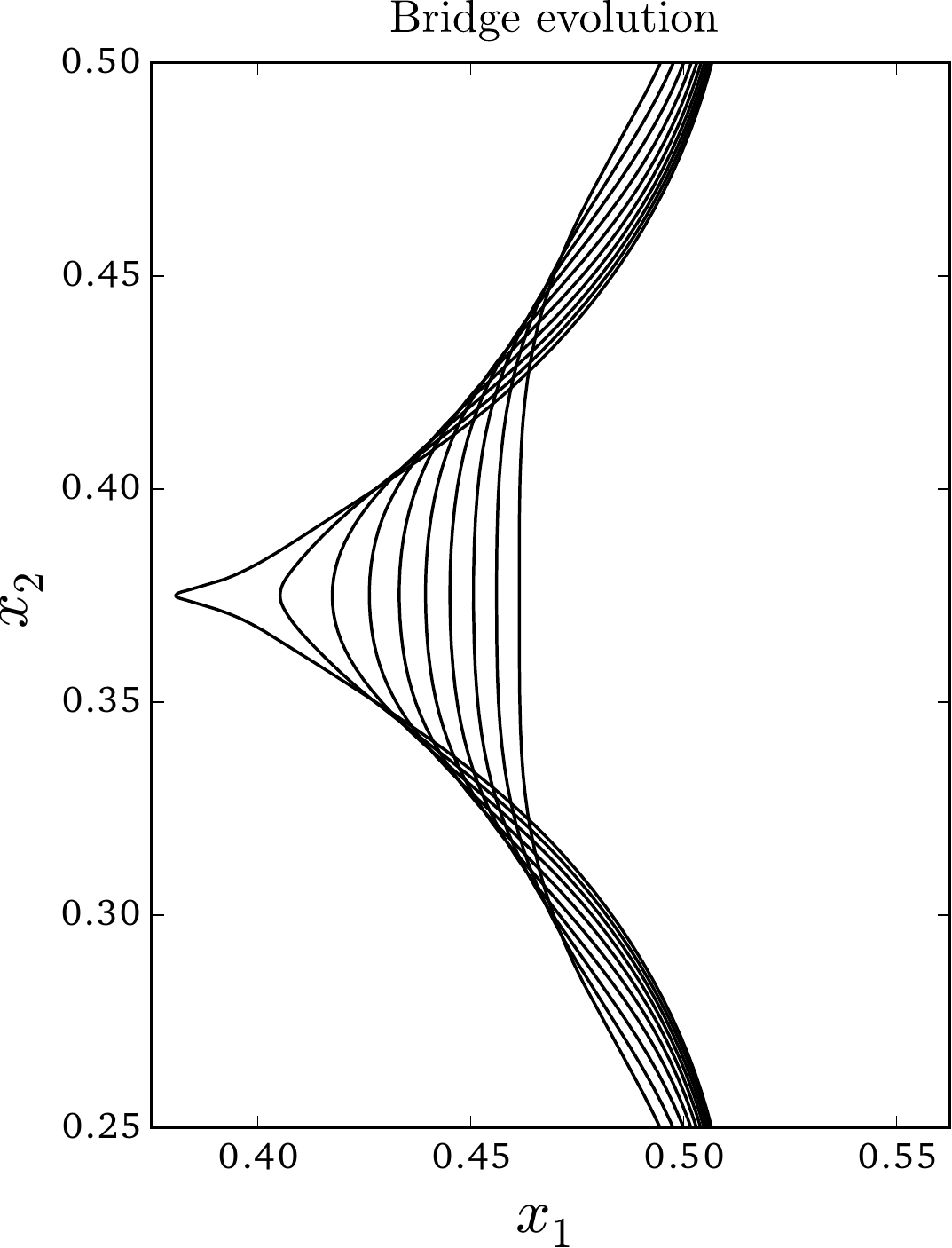}
  \caption{Visualization of the meniscus and bridge evolution on $ {\phi = 0.5} $. Case $ {\#1} $, from $ {t_i = 0.2698} $ to $ {t_f = t_i + 0.055} $ using equally spaced isocurves on time.}
\label{fg:meniscus.bridge.case.1}
\end{figure}

\begin{figure}
\centering
  \includegraphics[height=0.2\textheight]{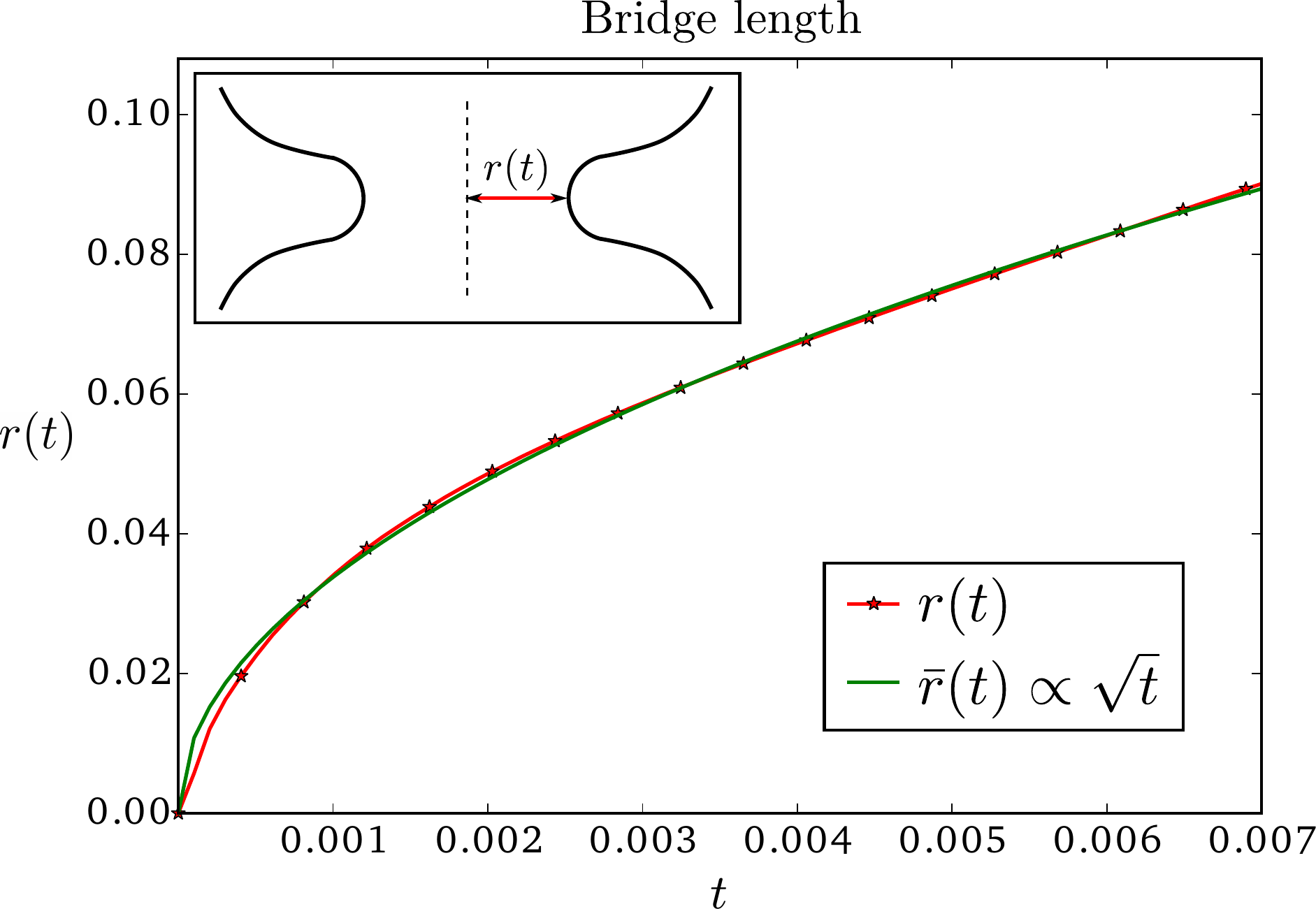}
  \caption{(color online) Bridge length evolution $ {r(t)} $, case $ {\#1} $. Red line - numerical result obtained herein, $ {r(t)} $. Green line - fitted square root curve, $ {\bar{r}(t) \propto \sqrt{t}} $.}
\label{fg:bridge.case.1}
\end{figure}

Figure \ref{fg:meniscus.bridge.case.1} depicts the meniscus evolution defined on the isocurve $ {\phi = 0.5} $ during the merging. Figure \ref{fg:bridge.case.1} shows the bridge length evolution, measured from the symmetry axis to the meniscus point. Due to the scaling law for the early stage, $ {r \lesssim 0.03 R} $, holds in a very tiny period of time, we restrict ourselves to analyze the bridge length for $ {r \gtrsim 0.03 R} $. Thus, we fit the bridge length obtained numerically, depicted by the red line in Figure \ref{fg:bridge.case.1}, with a fitted function $ {\sim \sqrt{t}} $, depicted by the green line in the same Figure.

The simulation results, obtained from the Navier--Stokes--Cahn--Hilliard equations, fit with the scaling law $ {\sim \sqrt{t}} $, also verified by the Stokes equation coupled to a sharp interface method \citep{EGG99} and by experimental results \citep{THO05,AAR05}. To further verify our model, we also fit the bridge length $ {r(t)} $ with a function $ {\sim t^b} $, being $ {b} $ a coefficient to be determined. The resulting coefficient is $ {b=0.504} $. As mentioned before, the theoretical prediction is $ {b=0.5} $ to yield the function $ {\sim \sqrt{t}} $. This allows us to conclude that there is an excellent agreement between our model and the theoretical/experimental models in the available literature.

\subsubsection{Energy Budget of the Flow: $ {\#1} $}

\begin{figure}[!t]
\centering
  \includegraphics[width=0.8\textwidth]{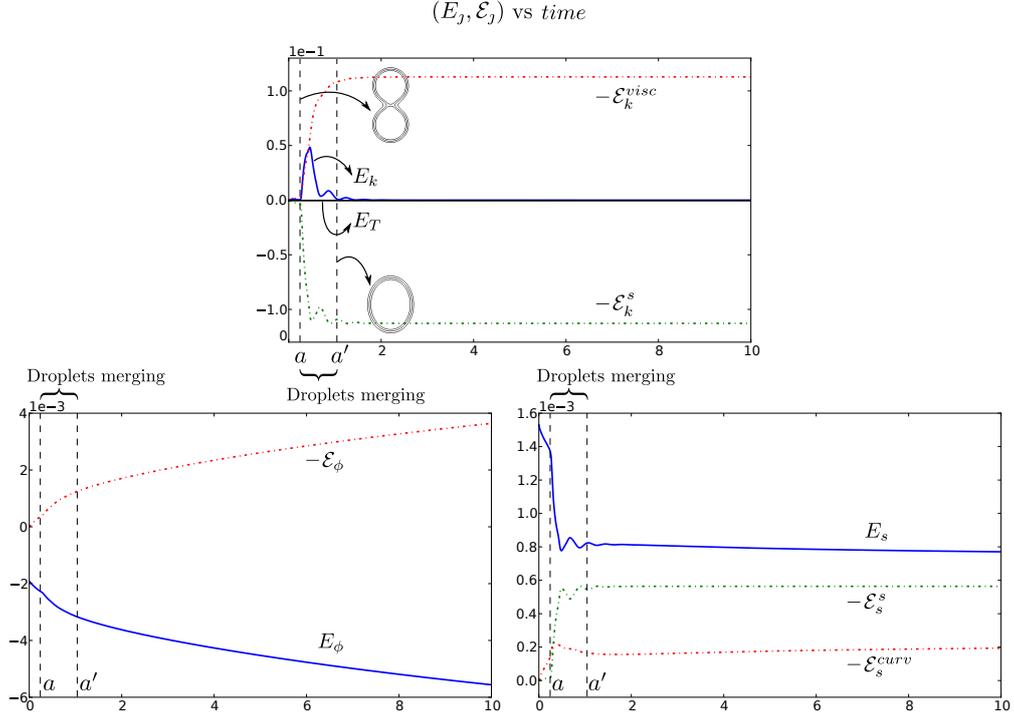}
\caption{(color online) Energy budget: case $ {\#1} $ 2D simulation. At the top: kinetic energy and its energy exchanges. At the bottom left: bulk free energy and its energy exchange. At the bottom right: interfacial free energy and its energy exchanges. Energies are depicted in solid lines and their energy exchanges in dashed lines.}
\label{fg:energy.budget.0}
\end{figure}

Figure \ref{fg:energy.budget.0} presents the extensive energies and their respective energy exchanges versus time, for case $ {\#1} $. At the top, the kinetic energy and its energy exchanges are shown. During the merging, the kinetic energy (solid blue line, $ {E_k} $) shows a peak which is quickly damped. The source of kinetic energy yielding that peak is provided by capillary effects and damped by the viscous ones. Energy exchange terms are depicted using dashed lines. The sink of the energy done by the viscous stress is depicted by the dashed red line, $ {- \mathcal{E}_k^{visc}} $, related to $ {\T{D}{2} : \T{T}{2}^{visc}} $, while the source of energy is provided by the capillary stress and depicted by the dashed green line, $ {- \mathcal{E}_k^{s}} $, related to $ {\T{D}{2} : \T{T}{2}^{s}} $. The term $ {- \mathcal{E}_k^{visc}} $ is always a sink of energy, thus its behavior is monotonic. In this experiment, the term $ {- \mathcal{E}_k^{s}} $ acts as a source of energy. Nevertheless, its behavior is non-monotonic, meaning that the overall behavior of the capillarity is not only a source but also a sink of energy.

At the bottom left, we depict the bulk free energy (solid blue line, $ {E_{\phi}} $) with its energy exchange term (dashed red line, $ {- \mathcal{E}_{\phi}} $), related to $ {\gradx{\eta_{\phi}} \cdot \T{j}{1}} $. The bulk free energy is decreasing and its energy exchange shows a monotonic increase. The overall behavior of this energy exchange is dissipative, i.e., it acts as a sink of energy. This means that the overall behavior of the mass flux has a positive inner product with the steepest descent direction of the phase field almost everywhere. This confirms our theoretical prediction (see \textbf{Remark \ref{rmk:mass.flux.0}}).

At the bottom right, we depict the interfacial free energy (solid blue line, $ {E_s} $) with its energy exchange terms (dashed red line, $ {- \mathcal{E}_s^{curv}} $) related to $ {\T{H}{2} : \gradx{\T{j}{1}}} $, and (dashed green line, $ {- \mathcal{E}_s^{s}} $) related to $ {\T{D}{2} : \gradx{\phi} \otimes \gradx{\phi}} $. The interfacial free energy decreases in a non-monotonic manner due to hydrodynamic effects, since the minimum surface area is not achieved monotonically. That is, the droplet oscillates around the circular configuration before settling at the steady configuration. A similar phenomenon is observed when an under-damped system oscillates around equilibrium. Both energy exchange terms appear as energy sinks. Regarding $ {- \mathcal{E}_s^{curv}} $, this means that the Hessian of $ {\phi} $, $ {\T{H}{2}} $, and the mass flux gradient, $ {\gradx{\T{j}{1}}} $, have a positive inner product. Finally, the term $ {- \mathcal{E}_s^{s}} $ has the same meaning as $ {-(- \mathcal{E}_k^{s})} $.

The total energy should be constant. However due to numerical errors and dissipations introduced by the time integrator we observe a decrease of around $ {0.07\%} $ in the total energy. Mass though is preserved exactly to machine-precision.

\subsection*{\centering Two-Dimensional Investigation: Cases $ {\#2} $ and $ {\#3} $}

In these examples, a gravitational field induces a density-driven flow and thus the droplets we consider in a liquid continuum rise and merge into a single bigger droplet.

\subsubsection{Meniscus and Bridge Evolution: $ {\#2} $ and $ {\#3} $}

\begin{figure}[!t]
\centering
  \includegraphics[height=0.3\textheight]{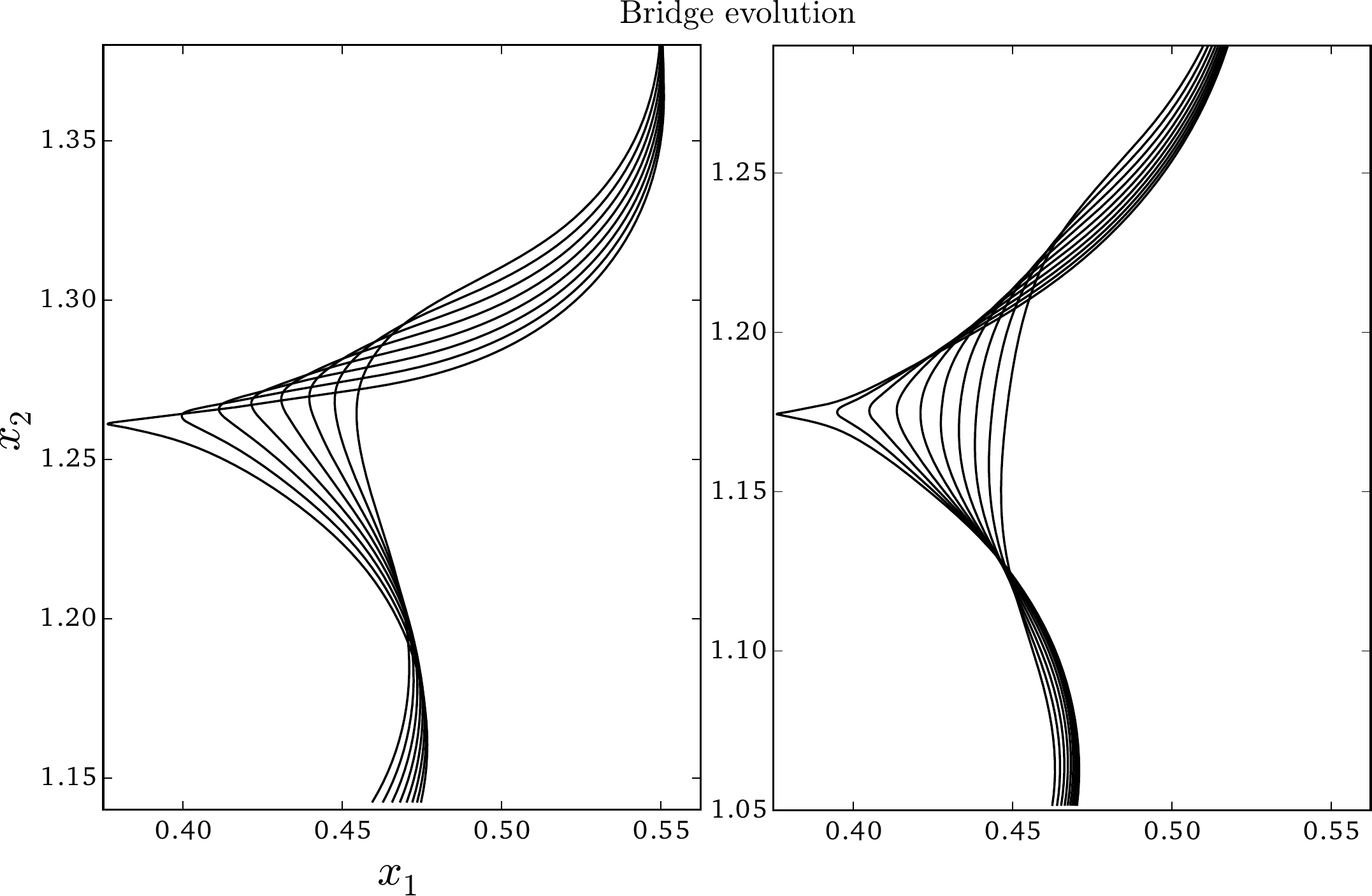}
  \caption{Visualization of the meniscus and bridge evolution on $ {\phi = 0.5} $. On the left, case $ {\#2} $, from $ {t_i = 4.12142} $ to $ {t_f = t_i + 0.04} $. On the right, case $ {\#3} $, from $ {t_i = 4.11107} $ to $ {t_f = t_i + 0.11} $. The isocurves are equally spaced on time.}
\label{fg:meniscus.bridge.case.2.3}
\end{figure}

\begin{figure}[!t]
\centering
  \includegraphics[height=0.2\textheight]{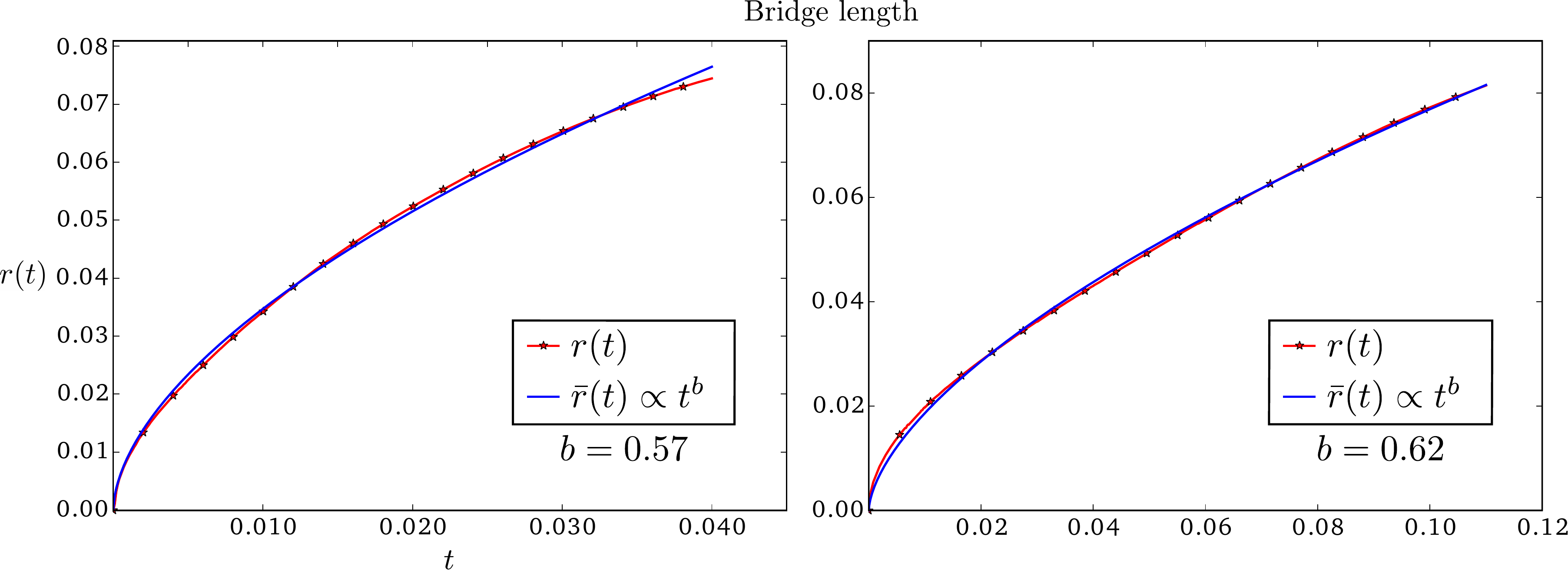}
  \caption{(color online) Bridge length evolution $ {r(t)} $ (red lines), case $ {\#2} $ and $ {\#3} $. Curves are fitted with a polynomial function $ {\bar{r}(t) \propto t^b} $.}
\label{fg:bridge.case.2.3}
\end{figure}

Since this experimental setup is beyond the assumptions of \citet{EGG99}, we fit the evolution of the bridge length with a function $ {\sim t^b} $. In Figure \ref{fg:meniscus.bridge.case.2.3} we depict the meniscus evolution defined on the isocurve $ {\phi = 0.5} $ during the rising and merging.
Figure \ref{fg:bridge.case.2.3} shows the bridge length evolution, measured from the symmetry axis to the meniscus point. The scaling law obtained by \citet{EGG99} ($ {\sim \sqrt{t}} $) does not hold for cases $ {\#2} $ and $ {\#3} $ due to buoyancy effects and different droplet radii. However, by using a least-squares fitting of the function $ {\sim t^b} $ for the exponent $ {b} $, we find that the bridge length can be approximated by $ {\sim t^{0.57}} $ and $ {\sim t^{0.62}} $ for cases $ {\#2} $ and $ {\#3} $, respectively. This suggests that the exponent $ {b} $ departs from $ {0.5} $ once buoyancy and non-uniform radii are considered. In case $ {\#1} $ the flow is driven by capillary effects, whereas in cases $ {\#2} $ and $ {\#3} $ the motion is mainly driven by buoyancy effects. This fact suggests that the flow may have different behaviors, i.e., different scaling laws.

\subsubsection{General Flow Features in Droplet Dynamics: $ {\#2} $ and $ {\#3} $}

\begin{figure}[!t]
\centering
  \includegraphics[width=0.55\textwidth]{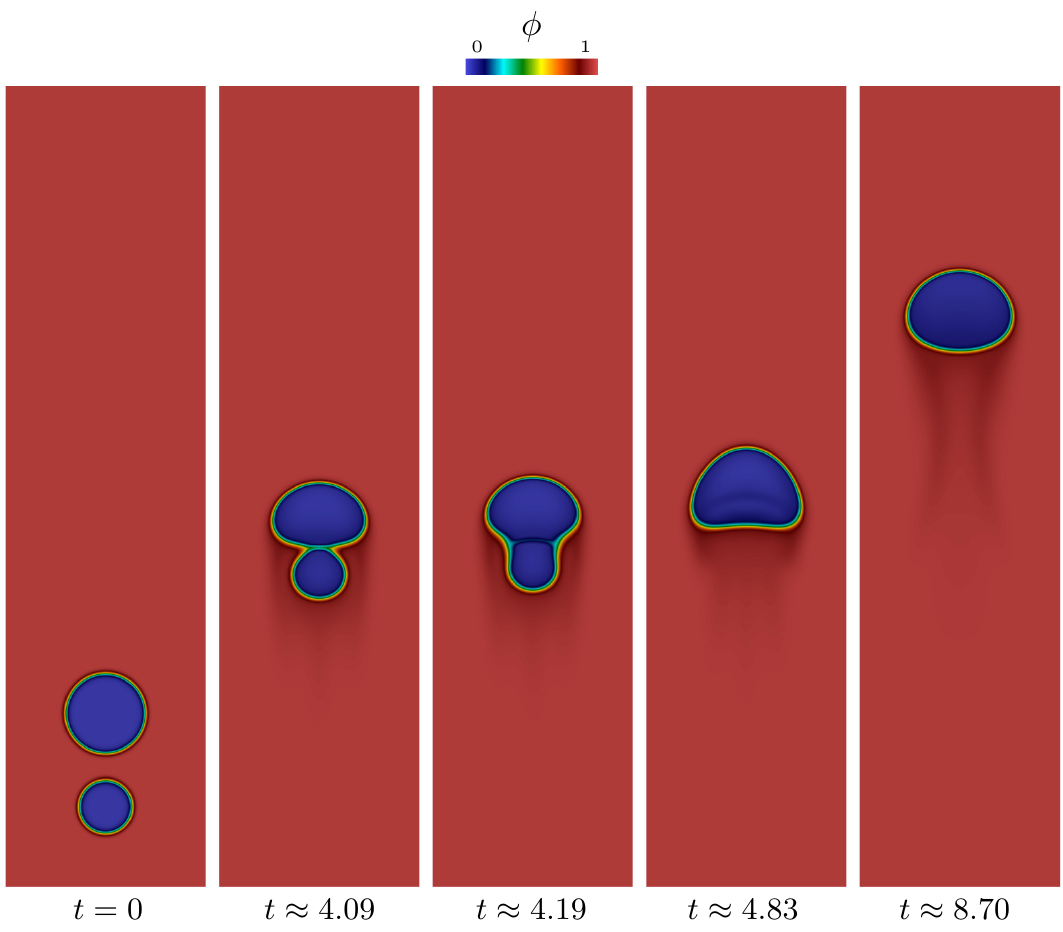}
\caption{(color online) Phase-field time evolution: case $ {\#2} $. From left to right the snapshots are related to times $ {t = 0} $, $ {\approx 4.09} $, $ {\approx 4.19} $, $ {\approx 4.83} $, and $ {\approx 8.70} $.}
\label{fg:phase.1}
% \end{figure}
\vspace{0.5cm}
% \begin{figure}
\centering
  \includegraphics[width=0.55\textwidth]{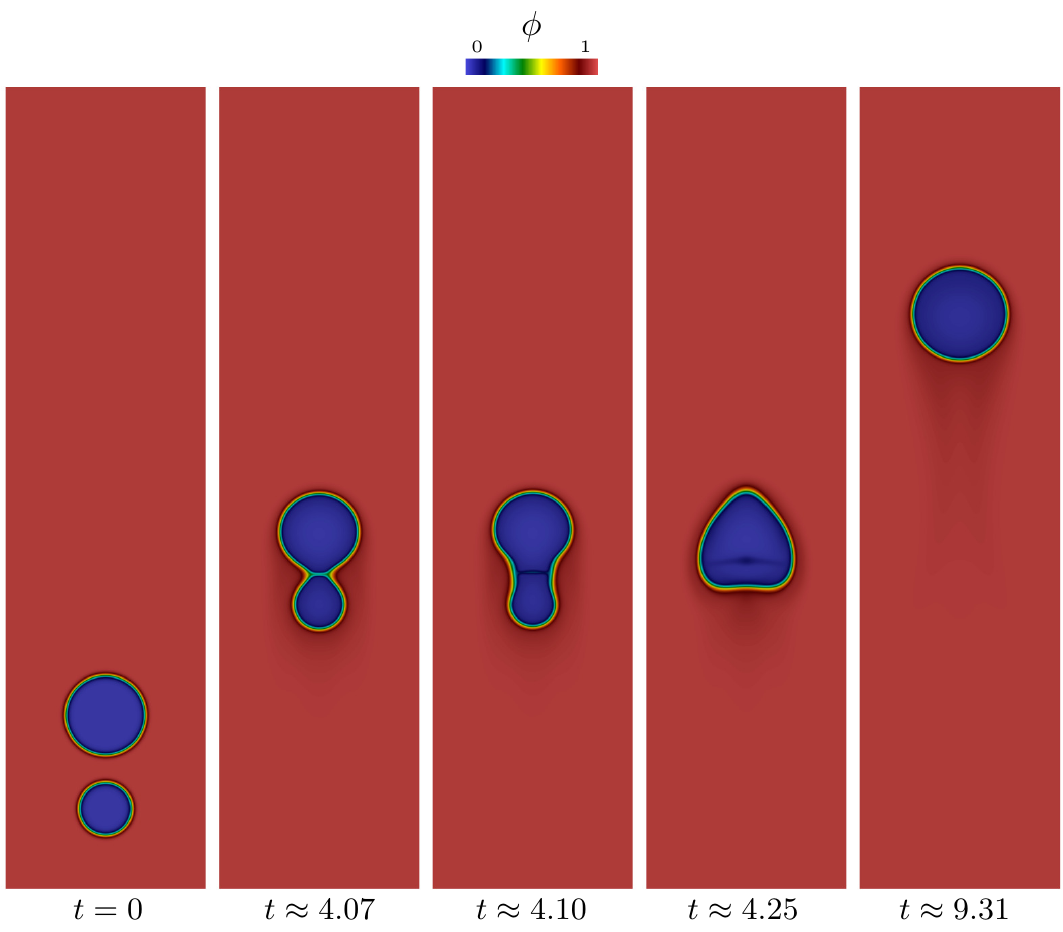}
\caption{(color online) Phase-field time evolution: case $ {\#3} $. From left to right the snapshots are related to times $ {t = 0} $, $ {\approx 4.07} $, $ {\approx 4.10} $, $ {\approx 4.25} $, and $ {\approx 9.31} $.}
\label{fg:phase.2}
\end{figure}

The temporal evolution of the phase field depicted in Figures \ref{fg:phase.1} and \ref{fg:phase.2} shows the phase-field dynamics during the merging and rising of droplets. In case $ {\#2} $, hydrodynamic effects lead to a variable interface thickness and a deformed shape as the droplets rise (cf. Figure \ref{fg:phase.1}). The leading zone shows a narrower interface, whereas the trailing zone shows a thicker one. The droplet obtained after the merging shows a double-elliptic shape, the lower ellipse being flatter than the upper one. Additionally, in case $ {\#3} $, as the $ {(Cn/We)^{-1}} $ ratio increases (cf. Figure \ref{fg:phase.2}), the surface tension dominates over the inertial effects, thus yielding a droplet which is almost spherical with constant interface thickness.

\begin{figure}[!t]
\centering
  \includegraphics[width=0.5\textwidth]{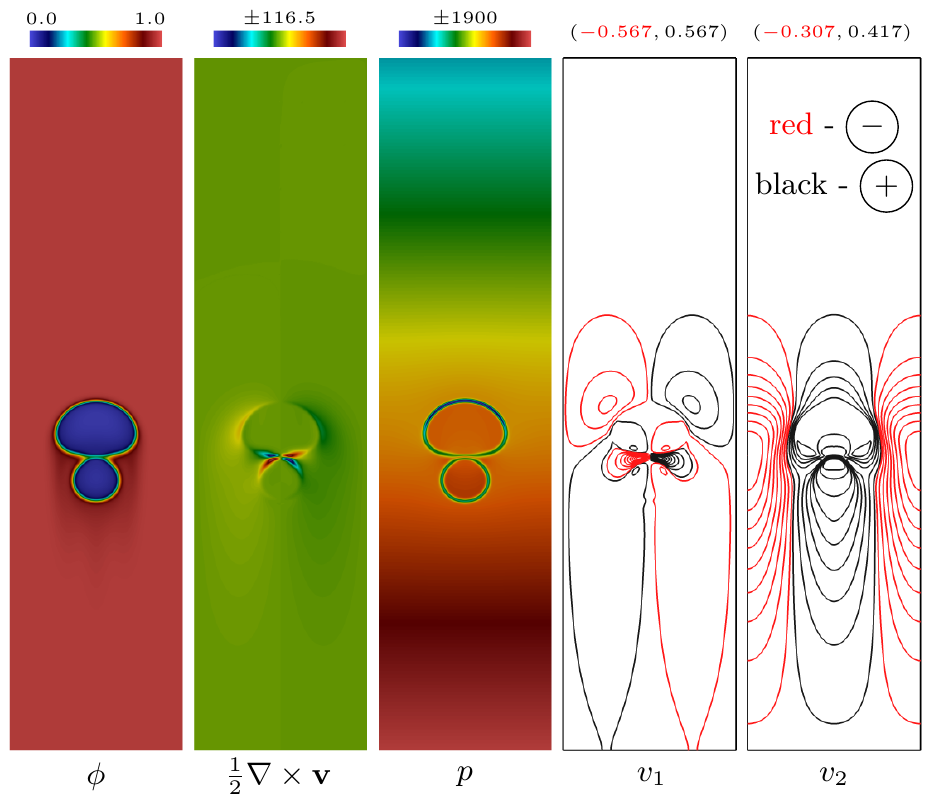}
\caption{(color online) Features of the flow: case $ {\#2} $. From left to right the snapshots of the phase field $ {\phi} $, vorticity $ {\frac{1}{2} \nabla\times \mathbf{v}} $, pressure $ {p} $, horizontal $ {v_1} $ and vertical $ {v_2} $ velocities are depicted at $ {t \approx 4.09} $.}
\label{fg:flow.1}
% \end{figure}
\vspace{0.5cm}
% \begin{figure}
\centering
  \includegraphics[width=0.5\textwidth]{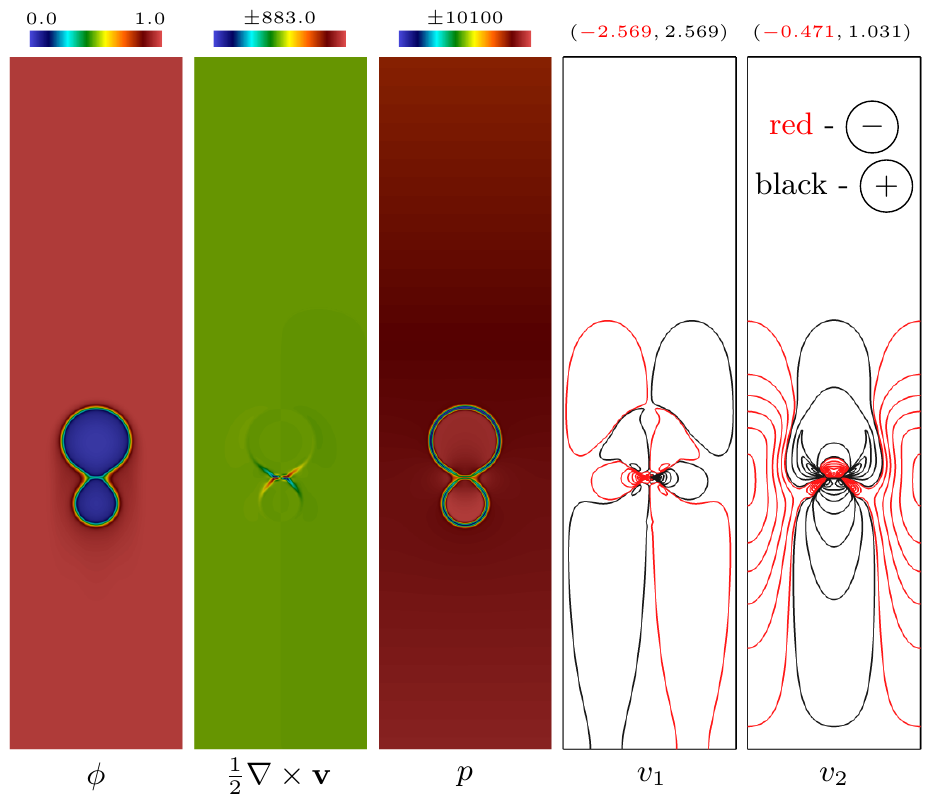}
\caption{(color online) Features of the flow: case $ {\#3} $. From left to right the snapshots of the phase field $ {\phi} $, vorticity $ {\frac{1}{2} \nabla\times \mathbf{v}} $, pressure $ {p} $, horizontal $ {v_1} $ and vertical $ {v_2} $ velocities are depicted at $ {t \approx 4.07} $.}
\label{fg:flow.2}
\end{figure}

To characterize the hydrodynamics; the phase, vorticity, pressure, and velocities are presented in Figures \ref{fg:flow.1} and \ref{fg:flow.2} for cases $ {\#2} $ and $ {\#3} $, respectively. Relevant flow features are shown when the merging of the droplets starts. The highest vorticity occurs in the meniscus zone. Case $ {\#3} $, with the highest $ {(Cn/We)^{-1}} $ ratio, shows a vorticity that is seven times larger than that of case $ {\#2} $. The pressure jump across the interface of the biggest droplet is an order of magnitude higher as $ {(Cn/We)^{-1}} $ grows by an order of magnitude. This jump grows from $ {10^3} $ to $ {10^4} $ as $ {(Cn/We)^{-1}} $ ratio grows from $ {10^2} $ to $ {10^3} $. Likewise, the velocities increase as $ {(Cn/We)^{-1}} $ ratio increases. The horizontal velocity, $ {v_1} $, increases by a factor of five, while the vertical velocity, $ {v_2} $, increases modestly by a factor of at most two.

The droplet merging process is essentially modified (from case $ {\#2} $ to $ {\#3} $) as the $ {(Cn/We)^{-1}} $ ratio increases by an order of magnitude. In case $ {\#2} $, the smaller (bottom) droplet pushes the larger (top) droplet, which accompanies the motion. In case $ {\#3} $, the droplets push against each other at the time of merging. The vertical velocities evidence these phenomena. In case $ {\#2} $, the vertical velocities inside the droplets are positive, whereas in case $ {\#3} $, the highest positive vertical velocity occurs in the northern hemisphere of the smaller (bottom) droplet and the highest negative vertical velocity is encountered at the southern hemisphere of the larger (top) one.

Stress profiles depicted in the last two snapshots of Figures \ref{fg:energies.1} and \ref{fg:energies.2} for cases $ {\#2} $ and $ {\#3} $, respectively, show the cross section integrals $ {\int T^{visc}_{22} \, \diffo{A_2}} $ and $ {\int T^{s}_{22} \, \diffo{A_2}} $ along the vertical axis $ x_2 $, with $ {\diffo{A_2} = \diffo{x_1} \diffo{x_3}} $. The capillary stress is higher at the interface, having its peak value at the leading zone during the merging and rising. The magnitude of the capillary stress is roughly five times higher for the highest $ {(Cn/We)^{-1}} $ ratio, case $ {\#3} $, also showing a smaller difference in the magnitude between the leading and trailing zones after merging (this difference decreases as the droplet takes a constant interface thickness). During the merging, the viscous stress is higher at the merging region for both cases. As the $ {(Cn/We)^{-1}} $ ratio increases, case $ {\#3} $, the viscous stress increases by a factor of two, whereas during the rising process this difference is the quintuple. During the rising, the resultant viscous stress takes positive values in the leading zone and negative ones in the trailing zone of the droplet. The local viscous stress is negative in the northern hemisphere of the droplet, but the viscous stress that lies outside of the droplet is positive and higher in magnitude. This is due to the higher viscosity of the interstitial fluid, which results in a positive viscous stress profile. The opposite behavior is observed in the southern hemisphere, which has a negative resultant viscous stress profile.

% \FloatBarrier

% \FloatBarrier

% \FloatBarrier

\subsubsection{Energy Exchange in the Merging and Rising of Droplets: $ {\#2} $ and $ {\#3} $}

\begin{figure}[!t]
\centering
  \includegraphics[width=0.55\textwidth]{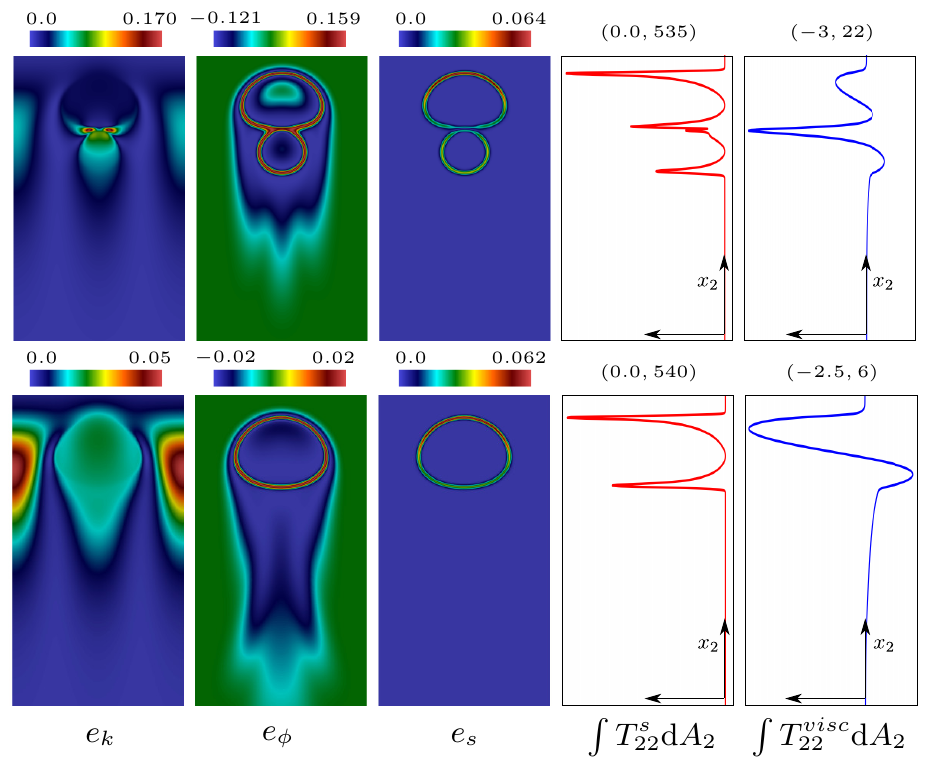}
\caption{(color online) Intensive energies and stress profiles at $ {t \approx 4.09} $ and $ {t \approx 8.70} $ at the top and at the bottom, respectively: case $ {\#2} $. From left to right: intensive kinetic energy $ {e_k} $, bulk free energy $ {e_{\phi}} $, interfacial free energy $ {e_s} $, capillary stress profile $ {T^{visc}_{22}} $ and viscous stress profile $ {T^{s}_{22}} $ ($ {\diffo{A_2} = \diffo{x_1} \diffo{x_3}} $).}
\label{fg:energies.1}
% \end{figure}
\vspace{0.5cm}
% \begin{figure}
\centering
  \includegraphics[width=0.55\textwidth]{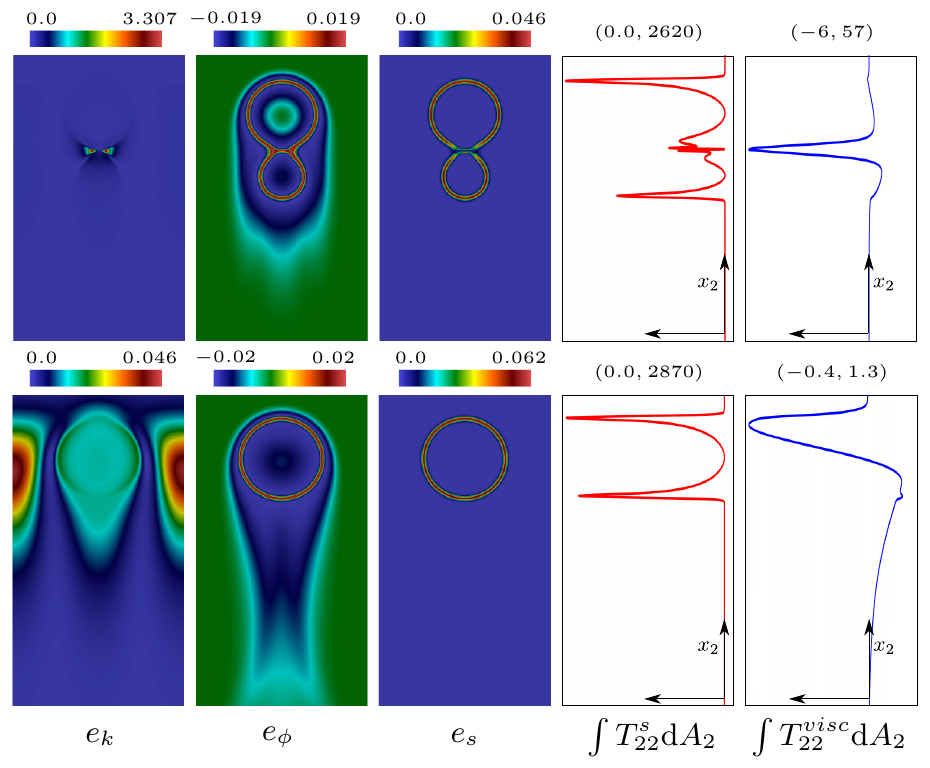}
\caption{(color online) Intensive energies and stress profiles at $ {t \approx 4.07} $ and $ {t \approx 9.31} $ at the top and at the bottom, respectively: case $ {\#3} $. From left to right: intensive kinetic energy $ {e_k} $, bulk free energy $ {e_{\phi}} $, interfacial free energy $ {e_s} $, capillary stress profile $ {T^{visc}_{22}} $ and viscous stress profile $ {T^{s}_{22}} $ ($ {\diffo{A_2} = \diffo{x_1} \diffo{x_3}} $).}
\label{fg:energies.2}
\end{figure}

\begin{figure}[!t]
\centering
  \includegraphics[width=0.5\textwidth]{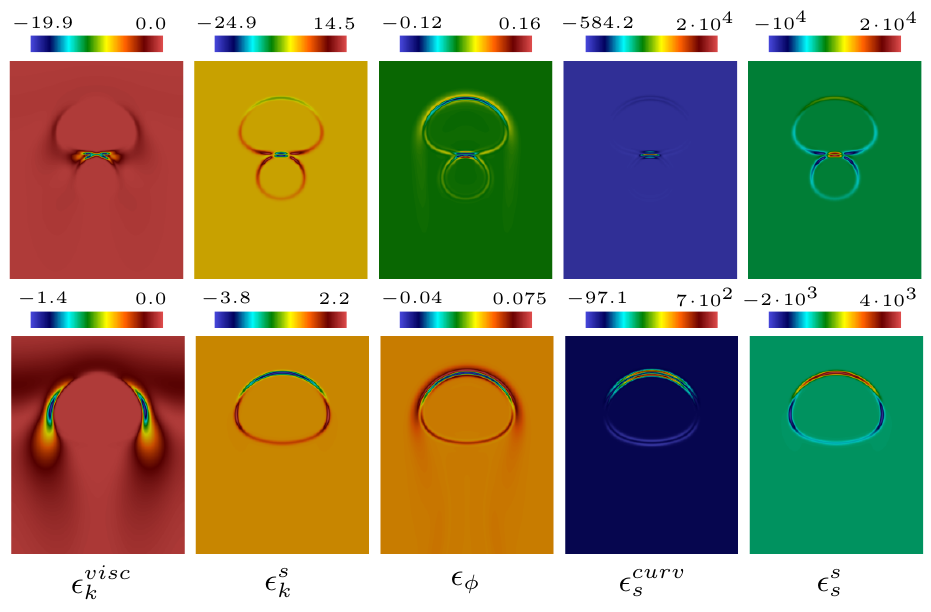}
\caption{(color online) Intensive energy exchanges at $ {t \approx 4.09} $ and $ {t \approx 8.70} $ at the top and at the bottom, respectively: case $ {\#2} $. From left to right: the intensive energy exchanges $ {\epsilon_k^{visc}} $, $ {\epsilon_k^{s}} $, $ {\epsilon_{\phi}} $, $ {\epsilon_{s}^{curv}} $ and $ {\epsilon_{s}^{s}} $.}
\label{fg:dissipations.1}
% \end{figure}
\vspace{0.5cm}
% \begin{figure}
\centering
  \includegraphics[width=0.5\textwidth]{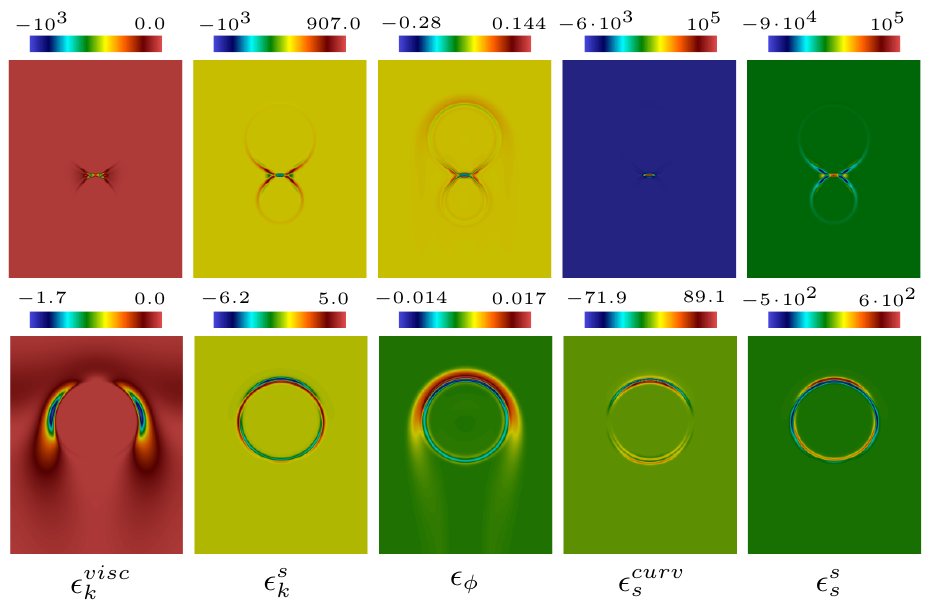}
\caption{(color online) Intensive energy exchanges at $ {t \approx 4.07} $ and $ {t \approx 9.31} $ at the top and at the bottom, respectively: case $ {\#3} $. From left to right: the intensive energy exchanges $ {\epsilon_k^{visc}} $, $ {\epsilon_k^{s}} $, $ {\epsilon_{\phi}} $, $ {\epsilon_{s}^{curv}} $ and $ {\epsilon_{s}^{s}} $.}
\label{fg:dissipations.2}
\end{figure}

During the merging of droplets, the highest intensive kinetic energy is located where the curvature of $ {\phi} $ is highest, i.e., in the meniscus region (see Figures \ref{fg:energies.1} and \ref{fg:energies.2} for cases $ {\#2} $ and $ {\#3} $, respectively). The source of this kinetic energy is given by the capillary stress. Evidence for this is taken from the second snapshot depicted in Figures \ref{fg:dissipations.1} and \ref{fg:dissipations.2} for cases $ {\#2} $ and $ {\#3} $, respectively, which correspond to the energy exchange $ {\epsilon_k^{s}} $. It is the source/sink given by the capillary stress $ {\T{D}{2} : \T{T}{2}^{s}} $ acting as a source of kinetic energy in the meniscus zone. Although the highest $ {(Cn/We)^{-1}} $ ratio, case $ {\#3} $, shows the highest kinetic energy (about 20 times greater than that observed for a lower $ {(Cn/We)^{-1}} $ ratio, case $ {\#2} $) after the merging, the resultant droplet in both cases presents similar values of intensive kinetic energy. This intensity decreases by one and two orders of magnitude after the merging phenomenon, for cases $ {\#2} $ and $ {\#3} $, respectively (cf. the second snapshot in Figures \ref{fg:energies.1} and \ref{fg:energies.2}).

\begin{figure}[!t]
\centering
  \includegraphics[width=0.5\textwidth]{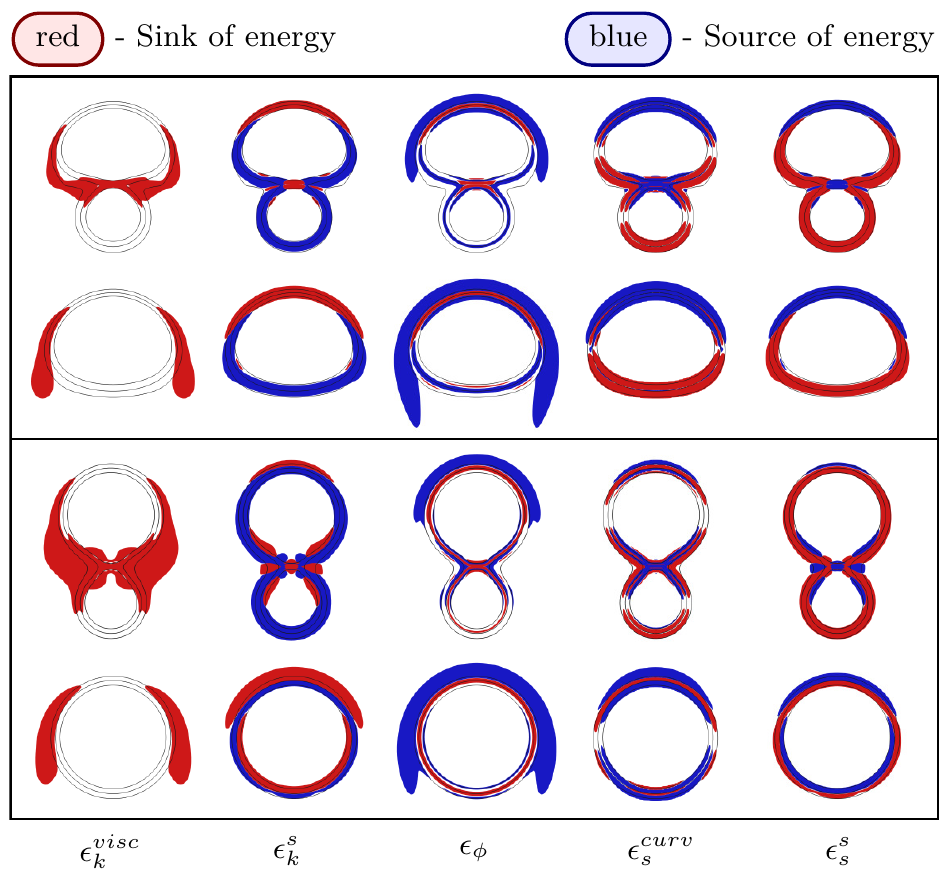}
\caption{(color online) Intensive energy exchanges, source (blue regions) and sink (red regions) of energy: case $ {\#2} $ at the top, $ {\#3} $ at the bottom.}
\label{fg:energy.regions}
\end{figure}

The viscous sink of kinetic energy $ {\T{D}{2} : \T{T}{2}^{visc}} $, i.e., the energy exchange $ {\epsilon_k^{visc}} $ (first snapshot depicted in \ref{fg:dissipations.1} and \ref{fg:dissipations.2} for cases $ {\#2} $ and $ {\#3} $, respectively) is also high in the confluence and meniscus zone. Increasing $ {(Cn/We)^{-1}} $ ratio by an order of magnitude ($ {\#2} \Rightarrow {\#3} $) yields a viscous sink greater by two orders of magnitude. Nonetheless, just before and after the merging, the region of highest viscous dissipation is located outside the droplet, in the interstitial fluid, near the equatorial region of the droplet (cf. Figure \ref{fg:energy.regions}), as usual in a flow around a sphere, with the same intensity in both cases. The capillary source/sink of kinetic energy $ {\T{D}{2} : \T{T}{2}^{s}} $ (second snapshot depicted in \ref{fg:dissipations.1} and \ref{fg:dissipations.2} for cases $ {\#2} $ and $ {\#3} $, respectively) acts as a sink of energy in the confluence region while in the meniscus region it acts as a source of energy. During the rising of the resultant droplet in the case of the lowest $ {(Cn/We)^{-1}} $ ratio, the capillary source of kinetic energy is located along the leading zone, whereas capillarity acts as a sink of kinetic energy on the trailing zone. However, in the case of the highest $ {(Cn/We)^{-1}} $ ratio, at the leading zone the capillarity acts as a source (sink) of kinetic energy just inside (outside) the droplet. At the trailing zone, the opposite behavior is observed.

In the merging of droplets, the intensity of the bulk free energy is greater by an order of magnitude for the lowest $ {(Cn/We)^{-1}} $ ratio, case $ {\#2} $. After the merging, the maximum and minimum values are equal for both cases. Meanwhile, the bulk free energy experiences a reduction by an order of magnitude for case $ {\#2} $ (lowest $ {(Cn/We)^{-1}} $ ratio), whereas for case $ {\#3} $ (highest $ {(Cn/We)^{-1}} $ ratio) the maximum and minimum values remain the same during the merging and rising. The bulk free energy density shows its highest value at the interface, whereas its lowest values define a hydrodynamic wake below the droplets as they rise (see the third snapshot in Figures \ref{fg:energies.1} and \ref{fg:energies.2} for cases $ {\#2} $ and $ {\#3} $, respectively).

The source/sink of bulk free energy occurs mainly at the interface (see the third snapshot in Figure \ref{fg:dissipations.1} and \ref{fg:dissipations.2} for cases $ {\#2} $ and $ {\#3} $, respectively). However, there is a region of energy production contouring the trailing zone that goes down towards the wake. This region arises due to hydrodynamic effects. In case $ {\#2} $ (lowest $ {(Cn/We)^{-1}} $ ratio) the bulk free energy exchange is a source and a sink at the trailing and leading zones, respectively. Nevertheless, in case $ {\#3} $ (highest $ {(Cn/We)^{-1}} $ ratio) the bulk free energy exchange is a sink at the interface.

As expected, the interfacial free energy is concentrated along the interface. During the merging, when the interfaces are coalescing, the interfacial energy decreases. In addition, at that time and place, the highest source of energy is provided by the first term in the interfacial energy exchange, $ {\epsilon_s^{curv} \propto \T{H}{2} : \gradx{\T{j}{1}}} $. The second term has the same effect as the capillary power for kinetic energy exchanges with the opposite sign.

To understand the overall instantaneous behavior of the energy exchanges, Figure \ref{fg:energy.regions} uses blue to indicate regions of energy sources and red to indicate regions of energy sinks.

% \FloatBarrier

% \FloatBarrier

% \FloatBarrier

\subsubsection{Energy Budget of the Flow: $ {\#2} $ and $ {\#3} $}

Figures \ref{fg:energy.budget.1} and \ref{fg:energy.budget.2} present the extensive energies and their related energy exchanges along time for cases $ {\#2} $ and $ {\#3} $, respectively. At the top left, the potential energy and its energy exchanges are shown. The potential energy (solid blue line, $ {E_p} $) shows a linear decay. Its first term (dashed red line, $ {- \mathcal{E}_p^{mass}} $) related to $ {\T{e}{1}_2 \cdot \T{j}{1}} $, in the energy exchange is almost zero. The second term (dashed green line, $ {- \mathcal{E}_p^{buoy}} $) related to $ {\phi \, v_2} $ in the potential energy exchange works as a sink of energy, however it cancels out with the third term of the kinetic energy exchange, $ {- \mathcal{E}_k^{buoy}} $.

At the top right, the kinetic energy and its energy exchanges are shown. During the merging, the kinetic energy (solid blue line, $ {E_k} $) shows a smooth (sharp) peak in case $ {\#2} $ (case $ {\#3} $) which is quickly damped. The energy exchange terms are depicted using dashed lines. The sink of energy done by the viscous stress is depicted by the dashed red line, $ {- \mathcal{E}_k^{visc}} $, related to $ {\T{D}{2} : \T{T}{2}^{visc}} $, while the sink (source) of energy in case $ {\#2} $ (case $ {\#3} $) is provided by the capillary stress and depicted by the dashed green line, $ {- \mathcal{E}_k^{s}} $, related to $ {\T{D}{2} : \T{T}{2}^{s}} $. The overall contribution $ {- \mathcal{E}_k^{s}} $ changes its behavior from a sink to a source of energy when the $ {(Cn/We)^{-1}} $ ratio increases. Nevertheless, in both cases the kinetic energy peak is provided by the energy exchange $ {- \mathcal{E}_k^{s}} $.

At the bottom left, the bulk free energy (solid blue line, $ {E_{\phi}} $) is depicted with its energy exchange term (dashed red line, $ {- \mathcal{E}_{\phi}} $) related to $ {\gradx{\eta_{\phi}} \cdot \T{j}{1}} $. In all examples presented here, the overall behavior of these energies is quite similar: the bulk free energy has a monotonic decay and its energy exchange acts as a sink of energy.

At the bottom right, the interfacial free energy (solid blue line, $ {E_{s}} $) is depicted with its energy exchange terms (dashed red line, $ {- \mathcal{E}_{s}^{curv}} $) related to $ {\T{H}{2} : \gradx{\T{j}{1}}} $, and (dashed green line, $ {- \mathcal{E}_{s}^{s}} $) related to $ {\T{D}{2} : \gradx{\phi} \otimes \gradx{\phi}} $. The interfacial free energy experiences a sudden drop during the merging. This drop increases as the $ {(Cn/We)^{-1}} $ ratio increases. Regarding $ {- \mathcal{E}_{s}^{curv}} $, this means that the Hessian of $ {\phi} $, $ {\T{H}{2}} $, and the mass flux gradient, $ {\gradx{\T{j}{1}}} $, have a positive inner product considering the overall behavior. Finally, the term $ {- \mathcal{E}_{s}^{s}} $ has the same meaning of $ {-(- \mathcal{E}_k^{s})} $.
\begin{figure}[!t]
\centering
  \subfloat[Case $ {\#2} $]{\label{fg:energy.budget.1}\includegraphics[width=0.8\textwidth]{energ1.pdf}} \\
% \end{figure}
\vspace{0.5cm}
% \begin{figure}
\centering
  \subfloat[Case $ {\#3} $]{\label{fg:energy.budget.2}\includegraphics[width=0.8\textwidth]{energ2.pdf}}
\caption{(color online) Energy budget: At the top left: potential energy and its energy exchange. At the top right: kinetic energy and its energy exchanges. At the bottom left: bulk free energy and its energy exchange. At the bottom right: interfacial free energy and its energy exchanges. Energies are depicted in solid lines and their energy exchanges in dashed lines.}
\label{fg:energy.budget.23}
\end{figure}

Due to numerical errors and dissipations introduced by the time integrator we observe a decrease less than $ {3.8\%} $ and $ {3.6\%} $ in the total energy, for cases $ {\#2} $ and $ {\#3} $, respectively. As in case $ {\#1} $, mass is preserved exactly to machine-precision.

\subsection*{\centering Three-Dimensional Investigation: Case $ {\#4} $}

In the three-dimensional case, the flow is also driven by buoyancy effects. The droplet is lighter and less viscous than the interstitial fluid.

\subsubsection{General Flow Features in Droplet Dynamics: $ {\#4} $}

\begin{figure}[!t]
\centering
  \includegraphics[width=0.7\textwidth]{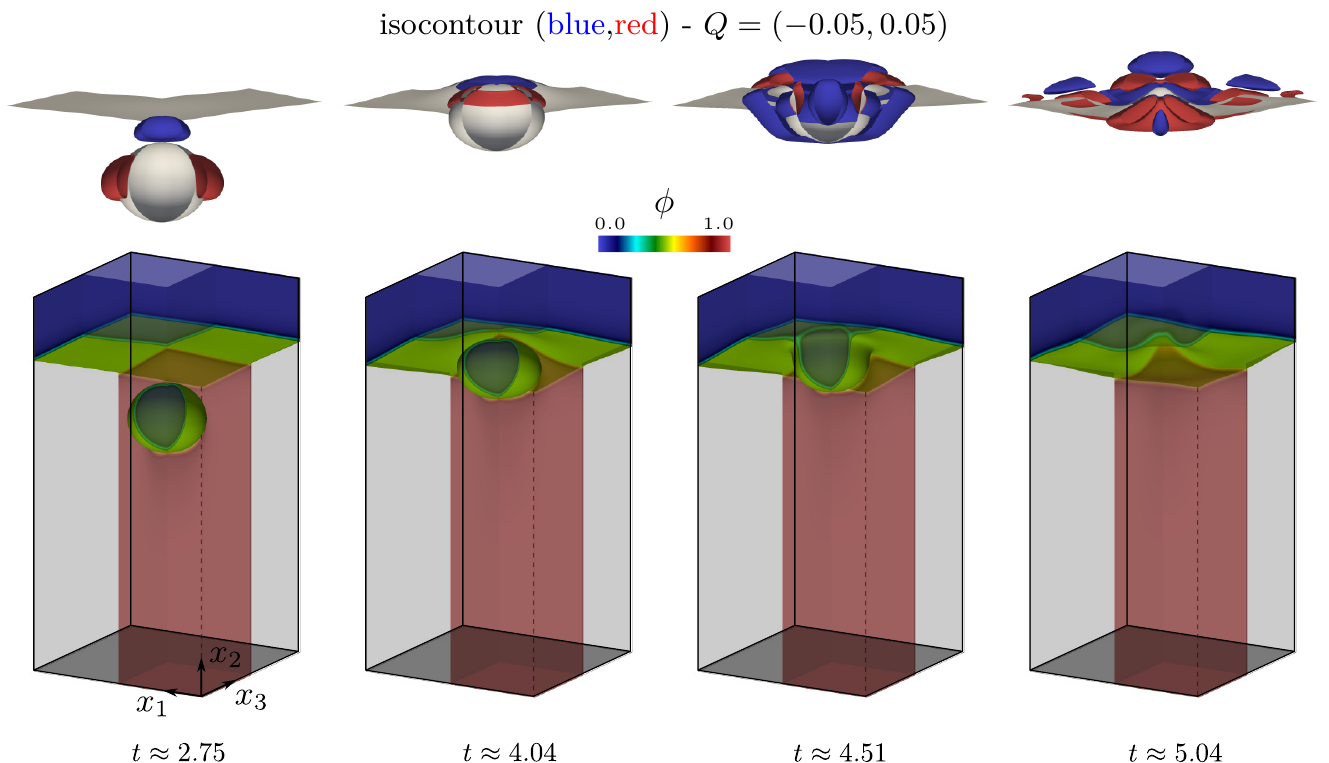}
\caption{(color online) At the top: Q-criterion evolution. At the bottom: phase-field evolution}
\label{fg:fields.3}
\end{figure}

The $ {Q} $-criterion is defined by the scalar $ {Q = \frac{1}{2} ( \T{W}{2} : \T{W}{2} - \T{D}{2} : \T{D}{2} )} $, where $ {\T{D}{2}} $ is the symmetric part of the velocity gradient, i.e., the strain rate tensor, and $ {\T{W}{2}} $ the anti-symmetric part of the velocity gradient, i.e., the spin rate tensor. Figure \ref{fg:fields.3} depicts at the top two isosurfaces of $ {Q} $-criterion, i.e., $ {Q = -0.05} $ and $ {0.05} $. Positive (negative) values show regions where the spin (strain) rate overcomes the strain (spin) rate.

\subsubsection{Energy Budget of the Flow: $ {\#4} $}

\begin{figure}[!t]
\centering
  \includegraphics[width=0.8\textwidth]{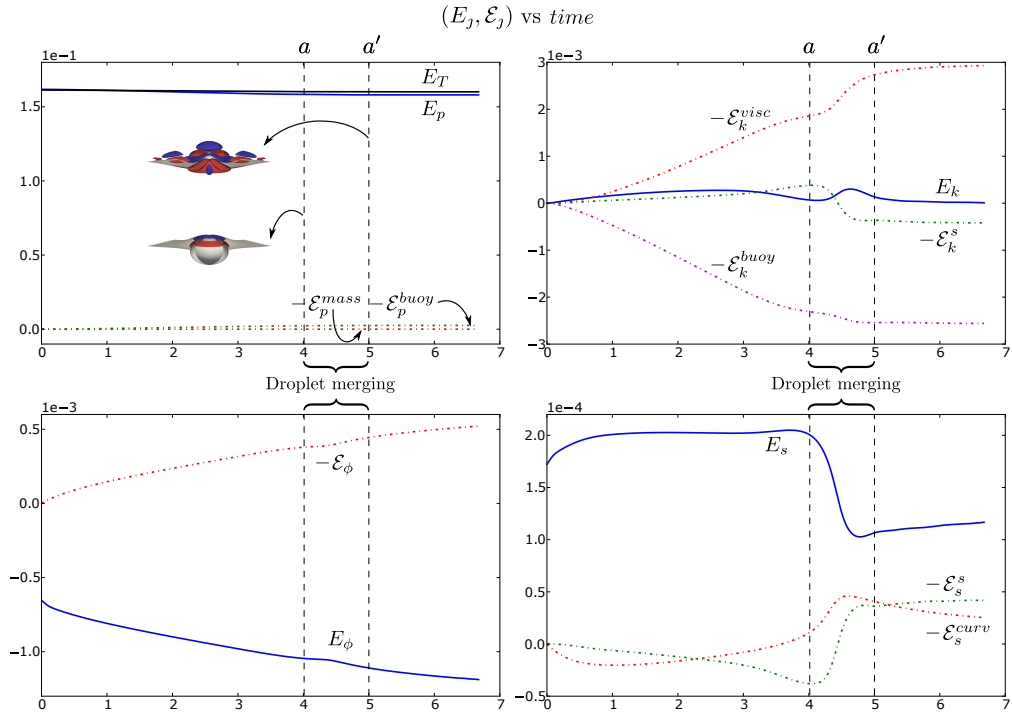}
\caption{(color online) Energy budget: case $ {\#4} $ 3D simulation. At the top left: potential energy and its energy exchange. At the top right: kinetic energy and its energy exchanges. At the bottom left: bulk free energy and its energy exchange. At the bottom right: interfacial free energy and its energy exchanges. Energies are depicted in solid lines and their energy exchanges in dashed lines.}
\label{fg:energy.budget.3}
\end{figure}

Figure \ref{fg:energy.budget.3} presents the extensive energies and their related energy exchanges along time, for this three-dimensional simulation, case $ {\#4} $. At the top left, the potential energy and its energy exchanges are shown. The potential energy (solid blue line, $ {E_p} $) shows a linear decay until the droplet merges in a thin film of fluid above the interstitial one, while its first term (dashed red line, $ {- \mathcal{E}_p^{mass}} $) related to $ {\T{e}{1}_2 \cdot \T{j}{1}} $ in the energy exchange, is almost zero. The second term (dashed green line, $ {- \mathcal{E}_p^{buoy}} $) related to $ {\phi v_2} $, works in the potential energy exchange as a sink of energy, but it cancels out with the third term of the kinetic energy exchange $ {- \mathcal{E}_k^{buoy}} $, as mentioned in cases $ {\#2} $ and $ {\#3} $.

At the top right, the kinetic energy and its energy exchanges are shown. Before the merging between the droplet and the thin film of fluid, the kinetic energy (solid blue line, $ {E_k} $) grows. When the droplet is close to the thin film the kinetic energy decreases. During the merging, the kinetic energy shows a smooth peak which is quickly damped by viscous effects. The energy exchange terms are depicted using dashed lines. The sink of energy contribution done by the viscous stress is depicted by the dashed red line, $ {- \mathcal{E}_k^{visc}} $, related to $ {\T{D}{2} : \T{T}{2}^{visc}} $, while the source/sink of energy provided by the capillary stress is depicted by the dashed green line, $ {- \mathcal{E}_k^{s}} $, related to $ {\T{D}{2} : \T{T}{2}^{s}} $. Before the merging process, capillary acts as a sink of energy. During the merging, capillary changes behavior to act as a source of energy.

At the bottom left, the bulk free energy (solid blue line, $ {E_{\phi}} $) is depicted with its energy exchange term (dashed red line, $ {- \mathcal{E}_{\phi}} $) related to $ {\gradx{\eta_{\phi}} \cdot \T{j}{1}} $. In all examples presented here, the overall behavior of these energies is quite similar.

At the bottom right, the interfacial free energy (solid blue line, $ {E_s} $) is depicted with its energy exchange terms (dashed red line, $ {- \mathcal{E}_s^{curv}} $) related to $ {\T{H}{2} : \gradx{\T{j}{1}}} $, and (dashed green line, $ {- \mathcal{E}_s^{s}} $) related to $ {\T{D}{2} : \gradx{\phi} \otimes \gradx{\phi}} $. Before the merging, the interfacial free energy grows to a stable baseline. However, the interfacial free energy experiences a sudden drop during the merging. The energy exchange term $ {- \mathcal{E}_s^{curv}} $ acts as a source of energy up to the merging, and changes its behavior during the merging to act as a sink of energy. A similar behavior is observed in the second term of the energy exchange, $ {- \mathcal{E}_s^{s}} $. Finally, the term $ {- \mathcal{E}_s^{s}} $ has the same meaning as $ {-(- \mathcal{E}_k^{s})} $.

Due to numerical errors and dissipations introduced by the time integrator, we observe a decrease less than $ {0.7\%} $ in the total energy. The error in the mass conservation is $ {0.002\%} $.

%===============================================================================================================%

\section{Conclusions}\label{sec:conclusions}

We develop the energy budget equation of the coupled Navier--Stokes--Cahn--Hilliard (NSCH) flow, the chief result of this work. We derive two new dimensionless numbers and denote these by $ {Lm} $ and $ {Ln} $, which relate the critical free energy density with the capillary energy, and the kinetic energy at the molecular scale with the critical free energy, respectively. In addition, we describe how the mass flux behaves in simple configurations, in one and two dimensions. We perform highly-resolved simulations, showing the energy exchanges and the energy budget of the flow. As a particular result, we show how the merging phenomenon of droplets is essentially modified when the $ {(Cn/We)^{-1}} $ ratio increases. To achieve such a detailed reproduction of the physical features of these multiphase flows we relied on robust and efficient (in-house and open-source) software tools, such as PetIGA and PetIGA-MF, to implement a robust discretization based on isogeometric analysis. We describe and analyze all energetic interactions at the discrete level as direct counterparts of those derived at the continuous level. We are able to achieve this by using divergence conforming discretizations for the velocity-pressure pair and a mixed form for the phase field and chemical potential. Finally, we show that our numerical simulations are in good agreement with the available analytical predictions and experimental result, concluding that employing NSCH equations to model droplet dynamics is a sensible approach worth of further research.

%===============================================================================================================%

\section{Acknowledgments}\label{sec:acknowledgments}

This work is part of the European Union's Horizon 2020 research and innovation programme of the Marie Sk{\l}odowska-Curie grant agreement No 644602.2014-0191. This publication also was made possible by a National Priorities Research Program grant NPRP grant 7-1482-1-278 from the Qatar National Research Fund (a member of The Qatar Foundation) and Agencia Nacional de Promoci\'on Cient\'{i}fica y Tecnol\'ogica (Argentina) grants \mbox{PICT~2014--2660} and \mbox{PICT-E~2014--0191}.

%===============================================================================================================%

\appendix

%===============================================================================================================%

\section{First and Second Laws of Thermodynamics}\label{sec:thermodynamics}

\subsection*{\centering First Law}

Here, we develop the first law in the classical manner, i.e., directly from the governing equations of momenta and mass transfer, cf. \citet{GUO15}. The inner product between the material derivative of the velocity and the velocity yields the balance between the external and internal mechanical rate of work done by the stress tensor. The product between the material derivative of the phase-field and the chemical potential yields the balance between the external and internal chemical rate of work done by the mass flux, including buoyancy effects.

We look for the following form
\begin{subequations}\label{eq:power.balance}
\begin{align}
\matd{e} _k + \matd{w} _m ^{int} = & \, \matd{w} _m ^{ext}, \label{eq:mech.power.balance} \\
\spart{\psi}{x_2}{} \, v_2 + \spart{\psi}{\phi}{} \, \matd{\phi} + \spart{\psi}{\gradx{\phi}}{} \cdot (\gradx{\phi}) \matd{} + \matd{w} _c ^{int} = & \, \matd{w} _c ^{ext}, \label{eq:chem.power.balance}
\end{align}
\end{subequations}
where $ {\matd{w} _m ^{int}} $ and $ {\matd{w} _m ^{ext}} $ ($ {\matd{w} _c ^{int}} $ and $ {\matd{w} _c ^{ext}} $) account for the internal and external mechanical power (internal and external chemical power), respectively. Here, $ {\spart{}{x_2}{}} $, $ {\spart{}{\phi}{}} $, and $ {\spart{}{\gradx{\phi}}{}} $ are the partial derivative with respect to $ {x_2} $, $ {\phi} $, and $ {{\gradx{\phi}}} $, respectively. Considering the free energy in the form (\ref{eq:ginzburg.landau.03}), i.e., $ {\psi = \psi _{\phi} + \psi _s + e _p} $. In the general case $ {\matd{\psi} = \spart{\psi}{x_2}{} \, v_2 + \spart{\psi}{\phi}{} \, \matd{\phi} + \spart{\psi}{\gradx{\phi}}{} \cdot (\gradx{\phi}) \matd{} + \spart{\psi}{\theta}{} \, \matd{\theta}} $, where $ {\spart{}{\theta}{}} $ is the partial derivative with respect to $ {\theta} $, whereas in the absence of heat transfer $ {\matd{\psi} = \spart{\psi}{x_2}{} \, v_2 + \spart{\psi}{\phi}{} \, \matd{\phi} + \spart{\psi}{\gradx{\phi}}{} \cdot (\gradx{\phi}) \matd{}} $. Thus, the Equation (\ref{eq:chem.power.balance}) is $ {\matd{\psi} + \matd{w} _c ^{int} = \matd{w} _c ^{ext}} $.

In the dimensional form, the internal and external powers are
\begin{subequations}\label{eq:powers}
\begin{align}
\matd{w} _m ^{int} = & \, \T{D}{2} : \T{T}{2}, \label{eq:powers.inte.m} \\
\matd{w} _m ^{ext} = & \, \divx{(\T{v}{1} \cdot \T{T}{2})} - \phi \jump{\rho} g v_2, \label{eq:powers.ext.m} \\
\matd{w} _c ^{int} = & \, - \gradx{\eta} \cdot \T{j}{1} + \gamma _{\phi} \T{D}{2} : \gradx{\phi} \otimes \gradx{\phi}, \label{eq:powers.int.c} \\
\matd{w} _c ^{ext} = & \, - \divx{(\eta \, \T{j}{1} - \gamma _{\phi} \matd{\phi} \gradx{\phi})} + \phi \jump{\rho} g v_2, \label{eq:powers.ext.c}
\end{align}
\end{subequations}
for a general framework, we may include a mass supply $ {j} $ in the right hand side in both Equations (\ref{eq:powers.ext.c}) and (\ref{eq:specie}).

The first law of thermodynamics represents an energy balance and states the interplay between the kinetic energy $ {e_k} $, the internal energy $ {e_i} $, the rate at which (mechanical and chemical) power is expended, and the rate at which energy in form of heat is transferred, i.e.,
\begin{equation}
\matd{e_T} = \matd{e} _k + \matd{e} _i = \matd{w} _m ^{ext} + \matd{w} _c ^{ext} - \divx{\T{q}{1}} + q,
\end{equation}
where $ {\T{q}{1}} $ is the heat flux, and $ {q} $ is a heat sink/source.

Finally, using the Equations (\ref{eq:powers}), we obtain the first law of thermodynamics
\begin{equation}\label{eq:first.law}
\begin{split}
\matd{e} _i = \spart{\psi}{x_2}{} \, v_2 + \spart{\psi}{\phi}{} \, \matd{\phi} + \spart{\psi}{\gradx{\phi}}{} \cdot (\gradx{\phi}) \matd{} + \T{D}{2} : \T{T}{2}^{visc} - \gradx{\eta} \cdot \T{j}{1} - \divx{\T{q}{1}} + q.
\end{split}
\end{equation}
In the absence of heat transfer the first thermodynamic law is
\begin{equation}\label{eq:first.law.no.heat}
\begin{split}
\matd{e} _i = \matd{\psi} + \T{D}{2} : \T{T}{2}^{visc} - \gradx{\eta} \cdot \T{j}{1}.
\end{split}
\end{equation}

\subsection*{\centering Second Law}

The second law of thermodynamics (in the form of the Clausius-Duhem inequality or entropy imbalance) states that the entropy $ {s} $ should grow at least with a rate given by the entropy flux $ {\T{q}{1}/\theta} $ added to the entropy supply $ {q/\theta} $, i.e.,
\begin{equation}\label{eq:entropy.growth}
\matd{s} \geqslant - \divx{\left(\dfrac{\T{q}{1}}{\theta}\right)} + \dfrac{q}{\theta} = \dfrac{1}{\theta} \left(- \divx{\T{q}{1} + \dfrac{1}{\theta} \gradx{\theta} \cdot \T{q}{1}} + q \right),
\end{equation}
where, $ {s} $ is the entropy. By definition, the free energy is
\begin{equation}
\psi = e _i - \theta s.
\end{equation}
Taking the material derivative, we obtain
\begin{equation}\label{eq:free.energy.time}
\matd{\psi} = \matd{e} _i - \matd{\theta} s - \theta \matd{s}.
\end{equation}

Replacing the first law (\ref{eq:first.law}) and Equation (\ref{eq:free.energy.time}) into Equation (\ref{eq:entropy.growth}), and considering that the entropy is $ {s = - \spart{\psi}{\theta}{}} $, the second law of thermodynamics is obtained in the form of the entropy imbalance, i.e.,
\begin{equation}\label{eq:entropy.imbalance}
\matd{s} = \dfrac{1}{\theta} \left( \T{D}{2} : \T{T}{2}^{visc} - \gradx{\eta} \cdot \T{j}{1} - \dfrac{1}{\theta} \gradx{\theta} \cdot \T{q}{1} \right) \geqslant 0.
\end{equation}
In the absence of heat transfer, the entropy imbalance yields
\begin{equation}\label{eq:entropy.imbalance.no.heat}
2 \mu (\phi) \T{D}{2} : \T{D}{2} + \alpha (\phi) \gradx{\eta} \cdot \gradx{\eta} \geqslant 0.
\end{equation}
Finally, this shows that our model guarantees the entropy production if $ {\mu (\phi), \alpha (\phi) \geqslant 0} $.

%===============================================================================================================%

\section{Identities used in the energy budget}\label{sec:identities}

To obtain the energy dissipation related to each energy, we employ some identities coupled with the governing equations as well as the constitutive relations. We include the ones used in our derivations here. It is worth noting that all identities has been derived considering that both second order tensors $ {\T{D}{2}} $ and $ {\T{H}{2}} $ are symmetric. 

Finally, in the kinetic energy
% \dfrac{\text{D}}{\text{D}t}(\dfrac{1}{2} \T{v}{1} \cdot \T{v}{1}) = & \, \T{v}{1} \cdot \dfrac{\text{D}\T{v}{1}}{\text{D}t} \\
% \divx{(\T{D}{2} \cdot \T{v}{1})} = & \, (\divx{\T{D}{2}}) \cdot \T{v}{1} + \T{D}{2} : \T{D}{2} \\
% \divx{(\gradx{\phi} \otimes \gradx{\phi}\cdot\T{v}{1})} = & \, \divx{(\gradx{\phi} \otimes \gradx{\phi})} \cdot \T{v}{1} + \T{D}{2} : \gradx{\phi} \otimes \gradx{\phi}
\begin{subequations}
\begin{align}
(\dfrac{1}{2} \T{v}{1} \cdot \T{v}{1}) \dot{} = & \, \T{v}{1} \cdot \dot{\T{v}{1}} \\
\divx{(\T{D}{2} \cdot \T{v}{1})} = & \, (\divx{\T{D}{2}}) \cdot \T{v}{1} + \T{D}{2} : \T{D}{2} \\
\divx{(\gradx{\phi} \otimes \gradx{\phi}\cdot\T{v}{1})} = & \, \divx{(\gradx{\phi} \otimes \gradx{\phi})} \cdot \T{v}{1} + \T{D}{2} : \gradx{\phi} \otimes \gradx{\phi}
\end{align}
\end{subequations}
In the bulk free energy
% \dfrac{\text{D}\psi_{\phi}(\phi)}{\text{D}t} = & \, \psi'_{\phi} \dfrac{\text{D}\phi}{\text{D}t} = \eta_{\phi} \dfrac{\text{D}\phi}{\text{D}t} \\
% \divx{( f(\phi) \, \T{H}{2} \cdot \gradx{\phi} )} = & \, f'(\phi) \, \T{H}{2} : \gradx{\phi} \otimes \gradx{\phi} \nonumber \\
% & + f(\phi) \, (\T{H}{2} : \T{H}{2} + (\divx{\T{H}{2}}) \cdot \gradx{\phi})
\begin{subequations}
\begin{align}
\dot{\psi}_{\phi}(\phi) = & \, \psi'_{\phi} \dot{\phi} = \eta_{\phi} \dot{\phi} \\
\divx{( f(\phi) \, \T{H}{2} \cdot \gradx{\phi} )} = & \, f'(\phi) \, \T{H}{2} : \gradx{\phi} \otimes \gradx{\phi} \nonumber \\
& + f(\phi) \, (\T{H}{2} : \T{H}{2} + (\divx{\T{H}{2}}) \cdot \gradx{\phi})
\end{align}
\end{subequations}
In the interfacial free energy
% \dfrac{\text{D}}{\text{D}t}(\dfrac{1}{2}\gradx{\phi}\cdot\gradx{\phi}) = & \, \gradx{\phi} \cdot \gradx{\dfrac{\text{D}\phi}{\text{D}t}} - \T{D}{2} : \gradx{\phi} \otimes \gradx{\phi} \\
% \gradx{\dfrac{\text{D}\phi}{\text{D}t}} = & \, \dfrac{\text{D}}{\text{D}t}(\gradx{\phi}) + \gradx{\T{v}{1}} \cdot \gradx{\phi} \\
% \divx{(\gradx{\phi} \cdot \gradx{\T{j}{1}})} = & \,\T{H}{2} : \gradx{\T{j}{1}} + \gradx{\phi} \cdot \Delta\T{j}{1} \\
% \divx{(f(\phi) \, \T{H}{2} \cdot (\divx{\T{H}{2}}))} = & \, f'(\phi) \, \T{H}{2} : \gradx{\phi} \otimes \divx{\T{H}{2}} \nonumber \\
% & + f(\phi) \, [ (\divx{\T{H}{2}}) \cdot (\divx{\T{H}{2}}) + \T{H}{2} : (\gradx{\divx{\T{H}{2}}})]
\begin{subequations}
\begin{align}
(\dfrac{1}{2}\gradx{\phi}\cdot\gradx{\phi}) \dot{} = & \, \gradx{\phi} \cdot \gradx{\dot{\phi}} - \T{D}{2} : \gradx{\phi} \otimes \gradx{\phi} \\
\gradx{\matd{\phi}} = & \, (\gradx{\phi}) \dot{} + \gradx{\T{v}{1}} \cdot \gradx{\phi} \\
\divx{(\gradx{\phi} \cdot \gradx{\T{j}{1}})} = & \,\T{H}{2} : \gradx{\T{j}{1}} + \gradx{\phi} \cdot \Delta\T{j}{1} \\
\divx{(f(\phi) \, \T{H}{2} \cdot (\divx{\T{H}{2}}))} = & \, f'(\phi) \, \T{H}{2} : \gradx{\phi} \otimes \divx{\T{H}{2}} \nonumber \\
& + f(\phi) \, [ (\divx{\T{H}{2}}) \cdot (\divx{\T{H}{2}}) + \T{H}{2} : (\gradx{\divx{\T{H}{2}}})]
\end{align}
\end{subequations}
In the potential energy
% \dfrac{\text{D}(\phi x_2)}{\text{D}t} = & x_2 \dfrac{\text{D}\phi}{\text{D}t} + \phi \, v_2 \\
% \divx{(x_2 \, \T{j}{1})} = & (\divx{\T{j}{1}}) \, x_2 + \T{j}{1} \cdot \T{e}{1}_2 \\
% \divx{(f(\phi) \, \T{e}{1}_2 \cdot \T{H}{2})} = & f'(\phi) \T{H}{2} : \gradx{\phi} \otimes \T{e}{1}_2 + f(\phi) \, \T{e}{1}_2 \cdot (\divx{\T{H}{2}})
\begin{subequations}
\begin{align}
(\phi x_2) \dot{} = & x_2 \dot{\phi} + \phi \, v_2 \\
\divx{(x_2 \, \T{j}{1})} = & (\divx{\T{j}{1}}) \, x_2 + \T{j}{1} \cdot \T{e}{1}_2 \\
\divx{(f(\phi) \, \T{e}{1}_2 \cdot \T{H}{2})} = & f'(\phi) \T{H}{2} : \gradx{\phi} \otimes \T{e}{1}_2 + f(\phi) \, \T{e}{1}_2 \cdot (\divx{\T{H}{2}})
\end{align}
\end{subequations}

%===============================================================================================================%

\bibliographystyle{elsarticle-harv}
%\bibliography{bib}

\end{document}